\documentclass[11pt]{article}

\usepackage[table,xcdraw,dvipsnames]{xcolor}
\usepackage{fullpage, parskip}

\usepackage{algorithm, algpseudocode}
\algtext*{EndFor}
\algtext*{EndIf}
\algtext*{EndWhile}

\usepackage{tikz}
\usepackage{natbib}

\usepackage{amsmath, amsfonts, amssymb, amsthm, mathrsfs, thmtools}

\newtheorem{theorem}{Theorem}

\newtheorem{lemma}{Lemma}
\newtheorem{claim}{Claim}

\usepackage{graphicx, tablefootnote, hhline, wrapfig, float, etoolbox}
\usepackage{booktabs, subcaption, multicol, multirow, xspace, soul}
\usepackage[normalem]{ulem}
\usepackage[title]{appendix}
\usepackage[T1]{fontenc}
\usepackage{tcolorbox, multicol}
\usepackage{hyperref}
\hypersetup{
    colorlinks=true,
    urlcolor=blue,
    citecolor=blue
}

\newcommand{\ALG}{\mathrm{ALG}}
\newcommand{\OPT}{\mathrm{OPT}}

\newcommand{\E}{\mathbb{E}}

\newcommand{\R}{\mathbb{R}}

\newcommand{\Z}{\mathbb{Z}}

\newcommand{\loss}{\ell}
\newcommand{\feas}{\textup{\texttt{feas}}}
\newcommand{\dist}{d}
\newcommand{\class}{\mathbf{C}}

\title{Provably Small Portfolios for Multiobjective Optimization with Application to Subsidized Facility Location}
\author{Swati Gupta \\ MIT \and Jai Moondra\footnote{Corresponding author, email: \texttt{jmoondra3@gatech.edu}} \\ Georgia Tech \and Mohit Singh \\ Georgia Tech}
\date{}

\begin{document}

\maketitle

\footnotetext{A different version of this article appeared in the \emph{Proceedings of the 24th ACM Conference on Economics and Computation}, 2023, and can be accessed at \url{https://arxiv.org/abs/2211.14873}.}

\begin{abstract}
    Many multiobjective real-world problems, such as facility location and bus routing, become more complex when optimizing the priorities of multiple stakeholders. These are often modeled using infinite classes of objectives, e.g., $L_p$ norms over group distances induced by feasible solutions in a fixed domain. Traditionally, the literature has considered explicitly balancing `equity' (or min-max) and `efficiency' (or min-sum) objectives to capture this trade-off. However, the structure of solutions obtained by such modeling choices can be very different. Taking a solution-centric approach, we introduce the concept of provably small set of solutions $P$, called a {\it portfolio}, such that for every objective function $h(\cdot)$ in the given class $\mathbf{C}$, there exists some solution in $P$ which is an $\alpha$-approximation for $h(\cdot)$. Constructing such portfolios can help decision-makers understand the impact of balancing across multiple objectives.     
    
    Given a finite set of base objectives $h_1, \ldots, h_N$, we give provable algorithms for constructing portfolios for (1) the class of conic combinations $\class = \{\sum_{j \in [N]}\lambda_j h_j: \lambda \ge 0\}$ and for (2) any class $\class$ of functions that interpolates monotonically between the min-sum efficiency objective (i.e., $h_1 + \ldots + h_N$) and the min-max equity objective (i.e., $\max_{j \in [N]} h_j$). Examples of the latter are $L_p$ norms and top-$\ell$ norms. As an application, we study the Fair Subsidized Facility Location (FSFL) problem, motivated by the crisis of medical deserts caused due to pharmacy closures. FSFL allows subsidizing facilities in underserved areas using revenue from profitable locations. We develop a novel bicriteria approximation algorithm and show a significant reduction of medical deserts across states in the U.S.
\end{abstract}

\section{Introduction}\label{sec: introduction}

\begin{figure}[t]
    \begin{minipage}{0.54\textwidth}
        \centering
        \includegraphics[width=0.9\textwidth]{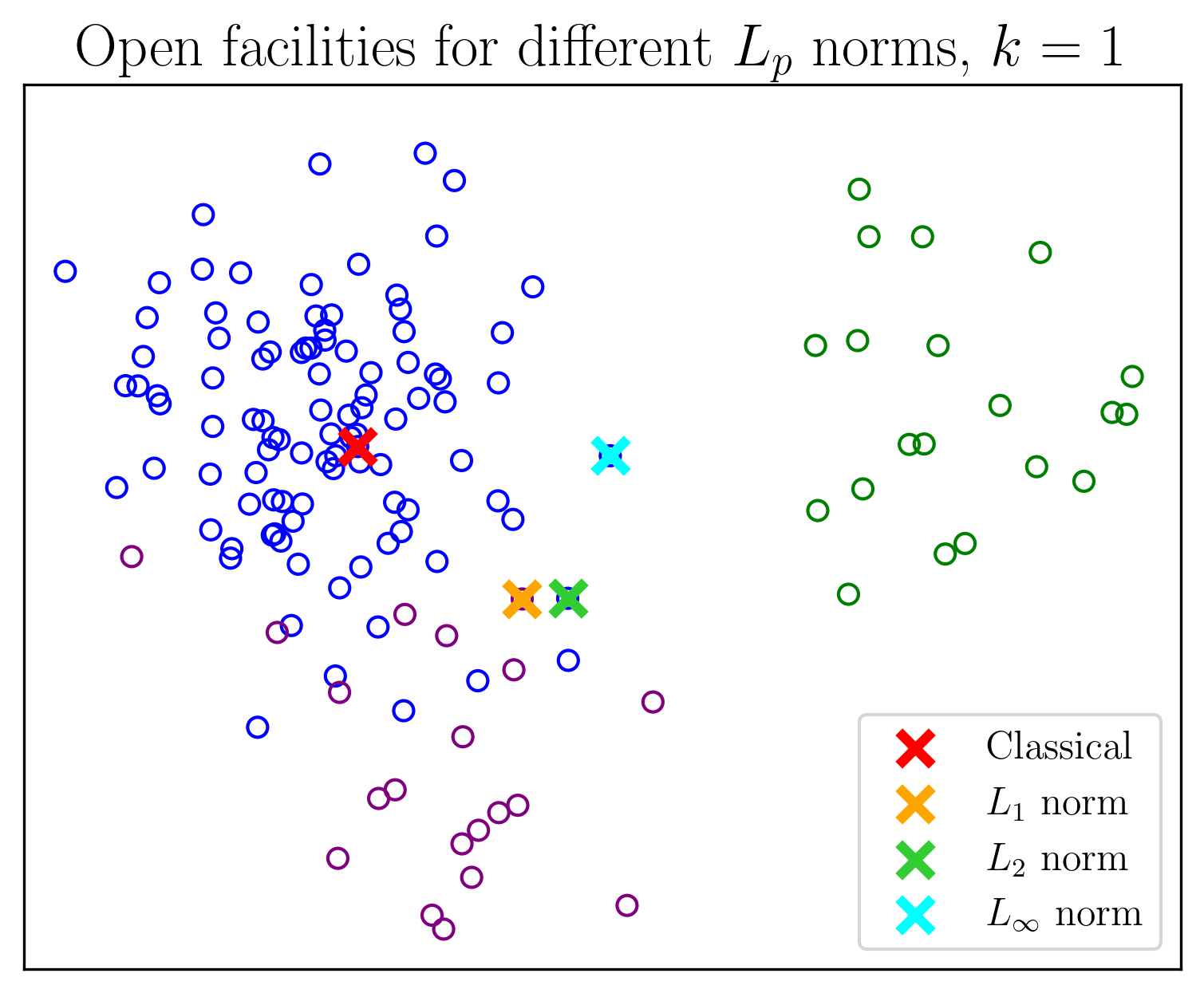}
        \footnotesize
        \begin{tabular}{|c|c|c|c|c|}
            \hline
            Group & Classical & $L_1$ norm & $L_2$ norm & $L_\infty$ norm \\ \hline
            1 (blue) & 0.21 & 0.42 & 0.47 & 0.46 \\ \hline
            2 (green) & 0.89 & 0.71 & 0.66 & 0.54 \\ \hline
            3 (purple) & 0.51 & 0.32 & 0.36 & 0.55 \\ \hline
            all clients & 0.35 & 0.45 & 0.48 & 0.48 \\ \hline
        \end{tabular}
    \end{minipage}
    \hfill
    \begin{minipage}{0.44\textwidth}
        \caption{An illustrative example for $k$-Clustering with three client groups $C_1, C_2, C_3$ (in blue, green, and purple, respectively) that partition client set $C$. We seek to open one facility $f$ anywhere in $C$. The optimal solution for the classical objective $\sum_{j \in C} \texttt{dist}_{j, f}$ opens facility $f$ near the center of the blue group. If we minimize the $L_p$ norm of vector $\left(\frac{1}{|C_s|} \sum_{j \in C_s} \texttt{dist}_{j, f}\right)_{s = 1, 2, 3}$ of average group distances, then $f$ moves closer to the center of all groups as $p$ increases from $1$ to $\infty$. The adjacent table shows average group distances for optimal solutions to different objectives.}
        \label{fig: three-groups-example}
    \end{minipage}
\end{figure}

The theory of multiobjective optimization \citep{evans_overview_1984,  masin_diversity_2008, grandoni_new_2014, tian_evolutionary_2021, herzel_approximation_2021, papadimitriou_multiobjective_2001} is now more critical than ever, as various applications require solutions that can achieve the right balance of efficiency, robustness to noise, stakeholder preferences, and equity. Instead of selecting a single composite objective function that explicitly balances these various objectives, we develop techniques to provide a small set of ``good enough'' solutions that provably approximate any function from a given set $\mathbf{C}$ of potential objectives of interest. For example, consider the problem of where to open pharmacies, with one objective being the cost of open facilities $h_1$, and the second objective being the distances traveled by different demographic groups $h_2$. There is a large class of objectives that one can consider to model $h_2$, e.g, the $L_p$ norm of the group distances vector $(d_1, \hdots, d_t)$ for $t$ groups, for some $p \ge 1$. One can ask to minimize the maximum distance traveled by any group (i.e., $L_{\infty}$ norm), or the total distance (i.e., the $L_1$ norm), or any $L_p$ norm with $p \in [1, \infty)$. Each of these choices yields a different objective, and hence a different optimal solution (see Figure \ref{fig: three-groups-example}). Traditional approaches might ask to fix a notion of equity (say $L_{\infty}$ norm) and fix a notion of efficiency (say $L_1$ norm), and find solutions at the Pareto-frontier of these two objectives. However, the bulk of existing work in this area (i) often does not give approximations for an arbitrary infinite set of objectives, and (ii) does not study the trade-off between solution quality and the size of the approximate Pareto frontier\footnote{The notion of a Pareto frontier for functions $h_1, \hdots, h_N$ requires a coordinate-wise non-dominance. However, the portfolios we introduce in this work can be much smaller in size due to guarantees on the composite function value (e.g., $\sum_i \lambda_i h_i(x) \leq \sum_i \lambda_i h_i(y)$ for all $y \in \mathcal{D}$). For more discussion, see Sections \ref{sec: results-conic-combinations} and \ref{sec: related-work}.}. We aim to alleviate these gaps in this work, and discuss connections to applications in transportation \citep{bertsimas_bus_2020, bertsimas_optimizing_2019}, finance \citep{fabozzi2008portfolio, papahristodoulou_optimal_2004}, clustering \citep{esmaeili_fair_2021, chierichetti_fair_2017}, etc. We further consider the specific problem of opening pharmacies to reduce medical deserts in the U.S., motivated by a recent study on the closure of pharmacies since 2015 \citep{NYT_Pharmacy_Deserts_2024}, and give new results for a novel model of the facility location problem with subsidies.  

\begin{figure}
    \centering
    \frame{\includegraphics[width=0.9\textwidth]{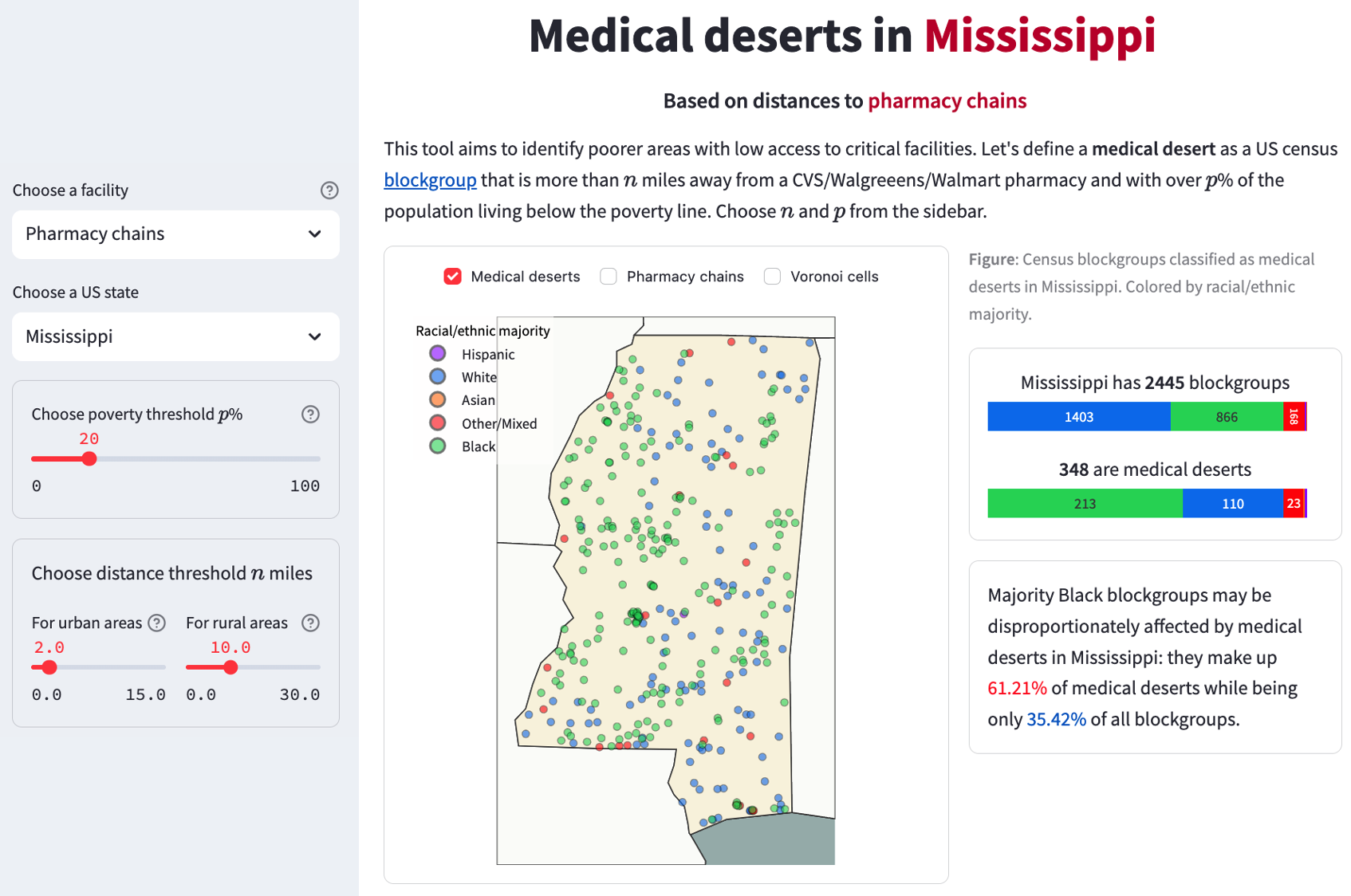}}
    \caption{A screenshot from our online tool depicting medical deserts in Mississippi, USA. Note that the majority Black blockgroups (35.42\% of all blockgroups) make up 61.03\% of all medical deserts. The tool can be accessed at \url{https://usa-medical-deserts.streamlit.app/}.}
    \label{fig: online-tool}
\end{figure}

Our key contribution is to introduce the concept of small portfolios of solutions for optimization problems, with provable guarantees with respect to all functions considered $\mathbf{C}$: 
\begin{center}
    \it Given an optimization problem with feasible solutions $\mathcal{D}$, a (potentially infinite) class $\mathbf{C}$ of objective functions and a desired approximation $\alpha \ge 1$, a set $P \subseteq \mathcal{D}$ is called an $\alpha$-approximate portfolio if for each  $h \in \mathbf{C}$, $\exists \ x\in P$ that is an $\alpha$-approximate solution to $h$, i.e., $h(x) \leq \alpha \min_{y\in \mathcal{D}} h(y)$.
\end{center}
If the portfolio size is restricted to 1, then this reduces to asking for a simultaneous approximation across all objectives (e.g., see \cite{KK00, goel_simultaneous_2006}). The flexibility of larger sizes allows us to study the trade-off between the portfolio size $|P|$ and the approximation factor $\alpha$. Next, we discuss the class of functions and applications of our framework.

\textbf{Class of functions.}
Given a finite set of ``base'' objective functions $h_1, \ldots, h_N: \mathcal{D} \to \R_{\ge 0}$ (e.g., access costs for groups $1$ to $N$), we consider the following canonical classes of functions: 
\begin{enumerate}
    \item[(i)] {\it Conic combinations of base functions}:  $\class_1 = \big\{\sum_{i \in [N]} \lambda_i h_i: \lambda \ge 0\big\}$, 
    \item[(ii)] {\it Interpolating functions}: $\class_2 = \{g_\lambda: \lambda \in [a, b]$ where $g_a(h_1, \ldots, h_N) = \|(h_1, \ldots, h_N)\|_1$, $g_b(h_1, \ldots, h_N) = \|(h_1, \ldots, h_N)\|_\infty\}$, which is any parametric class that interpolates monotonically between the egalitarian/equity (i.e., min max) and utilitarian/efficiency (i.e., min sum) objectives \citep{bentham1879principles, fleurbaey_theory_2010}.
\end{enumerate}

Special cases of $\class_2$ include the classes of (a) $L_p$ norm objectives $\big\{ \|(h_1, \ldots, h_N)\|_p: p\geq 1\big\}$ \citep{GGKT08}, (b) convex combinations of equity and efficiency objectives \citep{GJRYZ20, chen_just_2020}, and (c) top-$\ell$ norm objectives \citep{CS19}. We explore these in more detail in Section \ref{sec: monotonically-interpolating-portfolio}, and our results will apply to all these settings.

Beyond the foundational understanding of the concept of portfolios, we next shift our focus to applications. In particular, we next discuss a novel extension of the facility location problem motivated by the formation of medical deserts  \citep{NYT_Pharmacy_Deserts_2024} due to the closure of pharmacies. We show how to construct approximate oracles for this problem, which will then result in portfolios for facility location. We develop a web tool (see Figure \ref{fig: online-tool}) which can be useful for policymakers to guide investment decisions. We also discuss several other applications in Section \ref{sec: applications}.

\textbf{Fair Subsidized Facility Location.} Inequity in the placement of critical facilities is a well-documented problem \citep{multidimensional_poverty_2023, DPF12, calderon2004effects, GJRYZ20}.
For instance, it is suspected that profit maximization by grocery chains has led to the formation of food deserts spread widely across the U.S., which are defined as regions with low-income populations and low access to fresh food (e.g., people with no cars, and no grocery stores within a mile of their house) \citep{DPF12, balancing2022john}. Similarly, a recent New York Times study \citep{NYT_Pharmacy_Deserts_2024} showed that the closure of local pharmacies across the U.S. has resulted in the creation of pharmacy deserts. These smaller local stores often cannot sustain themselves in the wake of rising operating costs and drug prices, thus leading to closure. On the other hand, large pharmacy chains collect their revenues from a much broader source of users and have a higher negotiating power for drug prices, thus being more robust to financial difficulties.

We first consider a fair facility location problem, where given a $p \geq 1$, the goal is to minimize the $L_p$ norm of the distances traveled by each group of people. Our notion of portfolios is directly applicable here. We show that a portfolio of solutions can help us get around the ``choice'' of $p$, so the decision-maker can focus on the placement of facilities suggested by the portfolio, instead of debating modeling choices. Our model generalizes the extensively-studied classical $k$-Clustering and Uncapacitated Facility Location problems \citep{STA97, hochbaum1982heuristics, CG99, Li13}; see Theorem \ref{thm: clustering-and-ufl} in Appendix \ref{app: clustering-and-ufl}.

Specifically to optimize for profit-sustaining facilities, we propose a novel subsidized model of facility location where some of the profits from the large profit-making pharmacies can be diverted to the loss-making locations to sustain them, while improving access throughout all neighborhoods. Our model addresses the fundamental tension between {\it profitability} and {\it healthcare access}. Through the machinery of portfolios, we model how a central planner might strategically place pharmacies to ensure coverage in underserved areas, allowing some facilities to operate with losses subsidized by more profitable locations. These losses are bound by a fraction $\delta$ (called \emph{subsidy}) of the revenue that the planner can specify in advance, thus allowing them to be fairer to underserved communities while keeping losses limited. 

Formally, in Fair Subsidized Facility Location (FSFL), we are given a set of clients $C$ with non-negative revenues $r: C \to \R_{\ge 0}$, potential facilities $F$ with nonnegative operating costs $c: F \to \R_{\ge 0}$ and distances $\texttt{dist}$ on $C \cup F$ that form a metric, i.e., satisfy the triangle inequality. A solution consists of a set of open facilities $F^\prime \subseteq F$ and an assignment of each client to an open facility represented by $\Pi$. An open facility $f \in F'$ is profitable if its operating cost $c_f$ is less than the revenue $\sum_{j \in C: \Pi(j) = f} r_j$ brought by clients assigned to it, and unprofitable otherwise. The solution $(F', \Pi)$ is said to be $\delta$-\emph{subsidized} if the total loss of unprofitable facilities is at most a fraction $\delta$ of the total revenue $\sum_{j \in C} r_j$. Each client $j$ in the model can have a fractional\footnote{Fractional group memberships have not received much attention in the existing literature at the intersection of optimization, ML, and algorithmic fairness, which usually assumes that clients divide into non-intersecting groups, or considers the finest partition of intersections as unique individual groups \citep{DHPRZ12, chierichetti_fair_2017, celis2020interventions}.} group membership (e.g., dependent on geographic locations or socio-economic status) represented by $\mu_{j,s} \geq 0$ for each group $s \in [t]$.
The goal is to open a set $F' \subseteq F$ of facilities and provide an assignment $\Pi: C \to F'$ of clients to open facilities, so that (i) the solution is $\delta$-subsidized, and (ii) the distances of clients to facilities are ``equitable'' under various objectives, i.e., we would like to minimize the travel costs associated with $\class = \left\{\|\mathbf{h}^{(\Pi)}\|_p := \left(\sum_{s \in [t]} \big(h^{(\Pi)}_s\big)^p\right)^{1/p}: p \geq 1\right\},$ where $h^{(\Pi)}_s := \sum_{j \in C} \mu_{j, s} \texttt{dist}_{j, \Pi(j)}$ is the distance travelled by group $s \in [t]$. Note that opening fewer facilities leads to $\delta$-subsidized solutions, but this can increase access costs. Therefore, we would like to optimize access costs under various $p$-norms, while respecting that the solution is approximately $\delta$-subsidized.

\subsection{Technical Results}\label{sec: technical-results} Having discussed our modeling contributions, we next discuss our key technical results for (i) portfolios for conic combinations $\class_1$, (ii) portfolios for interpolating functions $\class_2$, and (iii) portfolios for FSFL. Finally, we end the section with a brief overview of our experimental results on constructing portfolios to recommend subsidized pharmacies in the U.S.

\subsubsection{Portfolios for Conic Combinations.}\label{sec: results-conic-combinations}
We first give portfolios for the class $\class = \{\sum_{j \in [N]} \lambda_j h_j: \lambda \in \R_{\ge 0}^{N}\}$ of conic combinations of given base objective functions, over any domain $\mathcal{D}$ such that each objective $\sum_{j} \lambda_j h_j$ can be optimized over $\mathcal{D}$ up to a $\beta$-approximation (for given $\beta \geq 1$, as we are considering minimization problems). The portfolio size is bounded in terms of the \emph{imbalance} $u:= \max_{i, j \in [N], x \in \mathcal{D}} \frac{h_i(x)}{h_j(x)}$ of the base objectives, defined as the maximum ratio between any two base functions at any point $x \in \mathcal{D}$ in the feasible set. Note that the base functions need not be convex.

\begin{restatable}{theorem}{convexCombinationsPortfolio}\label{thm: portfolios-convex-combinations}
    Let $h_1, \ldots, h_N: \mathcal{D} \to \mathbb{R}_{> 0}$ be positive (base) functions on feasible set $\mathcal{D}$ and some $u \in \mathbb{R}_+$ be such that at any point $x \in \mathcal{D}$ and any two functions $h_i, h_j$, it holds that $\frac{h_i(x)}{h_j(x)} \le u$. Let $\mathbf{C} = \{\sum_{i \in [N]} \lambda_i h_i: \lambda \in \R_{\ge 0}^N\}$ denote the class of all conic combinations of $h_1, \ldots, h_N$, and let $\mathcal{O}_\beta$ be an oracle to obtain a $\beta$-approximate solution to ${\arg\min}_{x \in \mathcal{D}} h(x)$ for any $h \in \class$ for given $\beta \ge 1$. Then, a $\beta(1 + \varepsilon)$-approximate portfolio of size  
        \[
            |P_\varepsilon| = N \cdot \left(\frac{12 \log\left(4Nu/\varepsilon \right)}{\varepsilon}\right)^{N - 1}
        \]
    can be constructed using at most $\textup{poly}(|P_\varepsilon|)$ number of oracle calls, for any $\varepsilon \in (0, 1]$.
\end{restatable}

For example, for base functions $h_1(x) = x+1, h_2(x) = x^2 + 1, h_3 = x^3-x^2+2$ over domain $\mathcal{D} =[0,1] \subseteq \R$, we have $u = 2$, and the above theorem gives a $(1 + \varepsilon)$-approximate portfolio of size $O\left(\frac{1}{\varepsilon^2} \left(\log\frac{1}{\varepsilon}\right)^2\right)$. Unlike previous results that use a multiplicative $\varepsilon$-mesh to construct such portfolios, our result combines it with an \emph{additive} $\varepsilon$-mesh, leading to better approximation ratios (see Sections \ref{sec: related-work} and \ref{sec: portfolios} for details). 
Despite the result's generality, the dependence on $N$ is exponential, but we show in Appendix \ref{app: convex-combination-portfolio-lower-bound} that this is unavoidable, unless there are more structural assumptions on $\class$ or the domain. We next present an example where a portfolio of a much smaller size can be constructed by exploiting the properties of the function class.

\emph{Example.} $\class = \{\sum_{i\in [N]} \lambda_i h_i: \lambda \ge 0\}$ on $\mathcal{D} = \R$, where $h_i(x) = \exp(\theta|x - y_i|)$, for $y_i$ uniformly sampled from $[0, 1]$ and some fixed $\theta > 0$. Then the grid $P = \left\{\frac{\log(1 + \varepsilon)}{\theta}, \frac{2 \log(1 + \varepsilon)}{\theta}, \ldots, 1 \right\}$ of $\frac{\theta}{\log(1 + \varepsilon)} \le \frac{2\theta}{\varepsilon}$ points is a $(1 + \varepsilon)$-approximate portfolio for any $\varepsilon \in (0, 1]$. 

In Example 1, note that the portfolio size of $2 \theta/\varepsilon$ is independent of $N$, and depends on the quality of the approximation factor and the growth of the functions in the class $\class$. We use similar ideas to generalize existing results (e.g., \cite{GGKT08}, \cite{goel_simultaneous_2006}) to show the existence of $(1 + \varepsilon)$-approximate portfolios of size $O\left(\frac{\log N}{\varepsilon}\right)$ for interpolating classes of objectives:

\begin{restatable}{theorem}{monotonicInterpolationPortfolio}\label{thm: portfolios-lp-norms}
    Let $h_1, \ldots, h_N: \mathcal{D} \to \mathbb{R}_{> 0}$ be $N$ positive functions on feasible set $\mathcal{D}$, and denote $\mathbf{h} = (h_1, \ldots, h_N)$. Let $\mathbf{C}$ denote a class of objectives that interpolate monotonically between $\|\mathbf{h}\|_1$ and $\|\mathbf{h}\|_\infty$, and let  $\mathcal{O}_\beta$ be an oracle that can give a $\beta$-approximate solution to ${\arg\min}_{x \in \mathcal{D}} h(x)$ for any $h \in \class$. There exists an algorithm that given any $\varepsilon \in (0, 1]$, finds a $\beta(1 + \varepsilon)$-approximate portfolio of size at most $O\left(\frac{\log \beta N}{\varepsilon}\right)$ in $\textup{poly}(N, \frac{1}{\varepsilon})$ number of oracle calls to $\mathcal{O}_\beta$.   
\end{restatable}

Note that Theorems \ref{thm: portfolios-convex-combinations} and \ref{thm: portfolios-lp-norms} use an approximation oracle $\mathcal{O}_\beta$ to construct the portfolio of solutions. Thus, they reduce the problem of finding portfolios to finding a $\beta$-approximate oracle for the underlying optimization problem for fixed objectives.

\textbf{Connections with the Pareto frontier.} Portfolios are closely connected to the notion of \emph{Pareto frontier} of base functions $h_1, \ldots, h_N$. Given points $x, y \in \mathcal{D}$, we say that $x \prec y$ or $x$ dominates $y$ if $h_i(x) \le h_i(y)$ for all $i \in [N]$ and strict inequality holds for some $i$. The Pareto frontier $F_{\mathbf{h}} = \{\mathbf{h}(x): x \in \mathcal{D} \text{\ is not dominated by any } y \in \mathcal{D}\}$ is the set of non-dominated function values. Given $\alpha \ge 1$, the $\alpha$-approximate Pareto frontier $F_{\mathbf{h}}(\alpha) := \left\{\mathbf{h}(x): x \in \mathcal{D}, \frac{1}{\alpha}\mathbf{h}(x) \in F_{\mathbf{h}} \right\}$ is the set of all function values that are coordinate-wise within an $\alpha$-approximation of the Pareto frontier.
Pareto frontiers require a coordinate-wise dominance, and in general can be much larger than portfolios for conic combinations of $h_1, \hdots, h_N$. For example, consider $\mathcal{D} = \{x \in \R^2: x_1 + x_2 = 1, x \ge 0\}$ and let $h_i$ simply be identity function for each coordinate $i$, i.e., $h_i(x) = x_i$ for $i \in \{1, 2\}, x \in \R^2$. Then, $\mathbf{h}(x) = x$ for all $x \in \R^2$. Since each point in $\mathcal{D}$ is non-dominated by any other, the Pareto frontier $F_\mathbf{h}$ must contain every point in $\mathcal{D}$. In particular, $F_\mathbf{h}$ is an infinite set. However, given any $\lambda_1,\lambda_2 \ge 0$, consider minimizing $\lambda_1 h_1(x) + \lambda_2 h_2(x) = \lambda_1 x_1 + \lambda_2 x_2$ over $\mathcal{D}$. Clearly, $(0, 1)$ or $(1, 0)$ is always optimal depending on whether $\lambda_1$ is greater than $\lambda_2$, and therefore, the set $\{(1, 0), (0, 1)\}$ of two points is an optimal portfolio for conic combinations of functions $h_1, h_2$.

\subsubsection{Bicriteria Oracle for the Facility Location Problem.} \label{sec: intro-fsfl}

We next discuss our contributions in designing an approximation oracle for Fair Subsidized Facility Location (FSFL) now. Like many other NP-hard problems \citep{byrka_bi-factor_2014, marathe_bicriteria_1998, grandoni_new_2014}, FSFL does not even admit a $\beta$-approximate oracle for any constant $\beta > 0$. In fact, even checking whether a given solution $x$ is feasible (i.e., whether $x \in \mathcal{D}$) is NP-hard for FSFL. We show this via a reduction from the Subset Sum problem (proof included in Appendix \ref{app: hardness}):

\begin{restatable}{theorem}{hardness}\label{thm: hardness}
    Unless P = NP, (A) FSFL is inapproximable to within any constant factor even when the objective is the sum of client distances, and (B) there is no polynomial-time algorithm to check the feasibility of a solution to FSFL.
\end{restatable}

In such cases, the natural next step is to relax one of the constraints by some factor $\gamma$ to obtain an extension $\mathcal{D}(\gamma) \supseteq  \mathcal{D}$ of the feasible set. For example, in the FSFL problem, one can relax the $\delta$-subsidy constraint to require that the total loss is at most $\gamma \times \delta \sum_j r_j$. An algorithm or oracle that returns solutions $x \in \mathcal{D}(\gamma)$ with objective value within factor $\alpha$ of the optimum in $\mathcal{D}$ is called a \emph{bicriteria $(\alpha, \gamma)$-approximation}.

Our definition of portfolios is general enough to accommodate bicriteria approximations:
{\it given an optimization problem with feasible solutions $\mathcal{D}$, a class $\mathbf{C}$ of objective functions and a desired $(\alpha, \gamma)$ bicriteria approximation, a \emph{portfolio} is a set $P \subseteq \mathcal{D}(\gamma)$ such that for each  $h \in \mathbf{C}$, $\exists \ x\in P$ so that it is an $(\alpha, \gamma$)-approximate solution to $h$, i.e.,  $h(x) \leq \alpha \min_{y\in \mathcal{D}} h(y),$ where $\mathcal{D}(\gamma)$ is the relaxation of $\mathcal{D}$.} 

The guarantees given in Theorems 1 and 2 also generalize to such bicriteria problems, as long as the required oracle can be constructed. In the case of FSFL, we allow the oracle to increase the total loss beyond fraction $\delta$ of the total revenue (while still bounding it), i.e., the $\delta$-subsidy condition. This extension is necessary for approximations and allows for meaningful trade-offs between profitability and access costs to facilities in our application. To state our result, we make the $\theta$-\emph{small revenues} assumption, wherein $\theta = \max_{j \in C, f \in F} \frac{r_j}{c_f}$ is defined as the maximum fraction of a facility's operating cost that can be met by a single client's revenue. Most commercial facilities like pharmacies rely on a large number of clients and so $\theta$ is typically very small. Our result formally states the following:

\begin{restatable}{theorem}{approximationAlgorithm}\label{thm: price-of-fairness}
    There exists a polynomial-time algorithm that given an instance of Fair Subsidized Facility Location (FSFL) that satisfies $\theta$-small revenues assumption and a subsidy $\delta > 0$, returns a $(2\delta + \theta$)-subsidized solution whose objective value is within factor $O\left(\max\left(1, \frac{1}{\delta}\right)\right)$ of the optimal $\delta$-subsidized solution.
\end{restatable}

Our algorithm builds on the Linear Programming rounding approximation algorithm of \cite{STA97} for $k$-Clustering and Uncapacitated Facility Location (UFL) with linear objectives, and our bound matches their approximations within constant factors. 
There are two challenges in extending their algorithm to FSFL: first, our objectives are sublinear rather than linear, and second, their algorithm does not account for the new subsidy constraint. We observe that their algorithm generalizes to sublinear objectives; however, the second challenge requires a new combinatorial subroutine (see Section \ref{sec: bounded-distances}) that ensures that the total losses are bounded. Moreover, there is no $\alpha$-approximation for UFL with linear objectives if $\alpha < 1.463$ unless P = NP \citep{sviridenko2002improved}, implying that our bound for $\delta > 1$ is tight up to constants. 

As a corollary of the above theorem, we obtain portfolios for FSFL with $t$ client groups:

\begin{restatable}{corollary}{portfolioFFL}\label{thm: portfolios-ffl-upper-bound}
    There exists a polynomial-time algorithm that given an instance of Fair Subsidized Facility Location (FSFL) with $t$ client groups that satisfies $\theta$-small revenues assumption and a subsidy parameter $\delta > 0$, obtains a portfolio $P$ of size $O\left(\log \left(\frac{t}{\min(1, \delta)}\right)\right)$ such that
    \begin{enumerate}
        \item each solution in the portfolio is $(2\delta + \theta)$-subsidized, and
        \item for each $p \ge 1$, there is a solution in the portfolio with objective value within factor $O\left(\max\left(1, \frac{1}{\delta}\right)\right)$ of the optimum $\delta$-subsidized solution for the $L_p$ norm objective.
    \end{enumerate}
\end{restatable}

In practice, this means that the algorithm can offer decision-makers a small menu of (say) 3-5 facility layouts that span a wide range of fairness preferences. This is computationally efficient and allows flexibility without overwhelming the user, as we later show in our experiments on US Census data and pharmacy chains like CVS, Walgreens, and Walmart. To complement the upper bound, we show the existence of instances of FSFL with $t$ client groups where any $O(1)$-approximate portfolio for $L_p$ norms must have size $\Omega(\log t)$. This shows that the portfolio size obtained by our algorithm is the best possible (up to constant factors):

\begin{restatable}{theorem}{portfolioFFLLowerBound}\label{thm: portfolios-ffl-lower-bound}
    There exist instances of FSFL with $\delta > 1$ where any $O(1)$-approximate portfolio for the class $\mathbf{C}$ of $L_p$ norms of groups distances with $t$ client groups must have size $\Omega(\log t)$.
\end{restatable}

Next, we complement our theoretical results with experiments on U.S. Census data and locations of pharmacy chains CVS, Walgreens, and Walmart.

\subsubsection{Experiments.} Recall our online tool (Figure \ref{fig: online-tool}) for identifying medical deserts: regions with over $20\%$ poverty rate and that are further than $n$ miles from their nearest CVS, Walgreens, or Walmart pharmacy, where $n = 2$ for urban areas and $n = 10$ for rural areas. In our experiments in this section, we propose 10 new pharmacies alongside 206 existing CVS, Walgreens, and Walmart pharmacies in the state of Mississippi, USA.
We divide the population into $t = 16$ groups based on the Congressional district, urbanization levels, and poverty levels, and give a portfolio of three solutions based on different $L_p$ norm objectives with these groups and for different subsidy parameters $\delta$.

Each solution in the portfolio recommends different facilities, thus \emph{offering a varied choice to the policymaker} (also see Table \ref{tab: spider-plots}). Despite their diversity, they all \emph{reduce the number of medical deserts identified by our tool from 348 to between 297-305 (depending on the solution)}, while opening only $10$ new facilities in all of Mississippi, and even with $\delta \le 0.02$, i.e., while losing only 2\% (additional\footnote{That is, the total loss of any \emph{new} facilities that we open must be within $2\%$ of the total revenue of all clients.}) revenue.
Further, $70$ to $80\%$ of these $\sim$50 blockgroups are majority Black or African American, thus \emph{mitigating some of the disproportionate impact of medical deserts on the Black population.}
These results are presented with further details in Section \ref{sec: experiments}. Our online tool, combined with these insights, could be a valuable resource for decision makers evaluating the locations of medical deserts and potential new pharmacy sites.

\begin{table}[]
\small
\begin{minipage}{0.53\linewidth}
\begin{tabular}{|ccc||llll|}
\hline
\multicolumn{3}{|c||}{Group} &
  \multicolumn{4}{c|}{Portfolio Solutions} \\ \hline
\multicolumn{1}{|c|}{\begin{tabular}[c]{@{}c@{}}Urban/\\ Rural\end{tabular}} &
  \multicolumn{1}{c|}{\begin{tabular}[c]{@{}c@{}}Poor/\\ Not Poor\end{tabular}} &
  \begin{tabular}[c]{@{}c@{}}Congress\\ District\end{tabular} &
  \multicolumn{1}{c|}{\begin{tabular}[c]{@{}c@{}}$L_1$\\ Norm\end{tabular}} &
  \multicolumn{1}{c|}{\begin{tabular}[c]{@{}c@{}}$L_{5.4}$\\ Norm\end{tabular}} &
  \multicolumn{1}{c|}{\begin{tabular}[c]{@{}c@{}}$L_{13.5}$\\ Norm\end{tabular}} &
  \multicolumn{1}{c|}{\begin{tabular}[c]{@{}c@{}}$L_\infty$\\ Norm\end{tabular}} \\ \hline
\multicolumn{1}{|c|}{} &
  \multicolumn{1}{c|}{} &
  1 &
  \multicolumn{1}{l|}{\cellcolor[HTML]{FCFCFF}0} &
  \multicolumn{1}{l|}{\cellcolor[HTML]{FCFCFF}0} &
  \multicolumn{1}{l|}{\cellcolor[HTML]{F8FBFC}\textbf{0.6}} &
  \cellcolor[HTML]{FCFCFF}0 \\ \cline{3-7} 
\multicolumn{1}{|c|}{} &
  \multicolumn{1}{c|}{} &
  2 &
  \multicolumn{1}{l|}{\cellcolor[HTML]{D1EBDA}6} &
  \multicolumn{1}{l|}{\cellcolor[HTML]{C4E5CE}7.8} &
  \multicolumn{1}{l|}{\cellcolor[HTML]{ACDCBA}\textbf{11}} &
  \cellcolor[HTML]{C1E5CD}8.1 \\ \cline{3-7} 
\multicolumn{1}{|c|}{} &
  \multicolumn{1}{c|}{} &
  3 &
  \multicolumn{1}{l|}{\cellcolor[HTML]{96D3A7}14} &
  \multicolumn{1}{l|}{\cellcolor[HTML]{8FD0A1}\textbf{15}} &
  \multicolumn{1}{l|}{\cellcolor[HTML]{8FD0A1}\textbf{15}} &
  \cellcolor[HTML]{9ED6AE}13 \\ \cline{3-7} 
\multicolumn{1}{|c|}{} &
  \multicolumn{1}{c|}{\multirow{-4}{*}{\begin{tabular}[c]{@{}c@{}}Not\\ Poor\end{tabular}}} &
  4 &
  \multicolumn{1}{l|}{\cellcolor[HTML]{DFF1E6}4} &
  \multicolumn{1}{l|}{\cellcolor[HTML]{DFF0E6}\textbf{4.1}} &
  \multicolumn{1}{l|}{\cellcolor[HTML]{F4F9F8}1.2} &
  \cellcolor[HTML]{F4F9F9}1.1 \\ \cline{2-7} 
\multicolumn{1}{|c|}{} &
  \multicolumn{1}{c|}{} &
  1 &
  \multicolumn{1}{l|}{\cellcolor[HTML]{FCFCFF}0} &
  \multicolumn{1}{l|}{\cellcolor[HTML]{FCFCFF}0} &
  \multicolumn{1}{l|}{\cellcolor[HTML]{FCFCFF}\textbf{0.12}} &
  \cellcolor[HTML]{FCFCFF}0 \\ \cline{3-7} 
\multicolumn{1}{|c|}{} &
  \multicolumn{1}{c|}{} &
  2 &
  \multicolumn{1}{l|}{\cellcolor[HTML]{A5D9B4}12} &
  \multicolumn{1}{l|}{\cellcolor[HTML]{79C78E}18} &
  \multicolumn{1}{l|}{\cellcolor[HTML]{79C78E}18} &
  \cellcolor[HTML]{72C488}\textbf{19} \\ \cline{3-7} 
\multicolumn{1}{|c|}{} &
  \multicolumn{1}{c|}{} &
  3 &
  \multicolumn{1}{l|}{\cellcolor[HTML]{63BE7B}\textbf{21}} &
  \multicolumn{1}{l|}{\cellcolor[HTML]{63BE7B}\textbf{21}} &
  \multicolumn{1}{l|}{\cellcolor[HTML]{63BE7B}\textbf{21}} &
  \cellcolor[HTML]{63BE7B}\textbf{21} \\ \cline{3-7} 
\multicolumn{1}{|c|}{\multirow{-8}{*}{Rural}} &
  \multicolumn{1}{c|}{\multirow{-4}{*}{Poor}} &
  4 &
  \multicolumn{1}{l|}{\cellcolor[HTML]{96D3A7}\textbf{14}} &
  \multicolumn{1}{l|}{\cellcolor[HTML]{B4DFC1}10} &
  \multicolumn{1}{l|}{\cellcolor[HTML]{F2F8F7}1.4} &
  \cellcolor[HTML]{F8FBFC}0.63 \\ \hline
\multicolumn{1}{|c|}{} &
  \multicolumn{1}{c|}{} &
  1 &
  \multicolumn{1}{l|}{\cellcolor[HTML]{FCFCFF}0} &
  \multicolumn{1}{l|}{\cellcolor[HTML]{FCFCFF}0} &
  \multicolumn{1}{l|}{\cellcolor[HTML]{FCFCFF}0} &
  \cellcolor[HTML]{FCFCFF}0 \\ \cline{3-7} 
\multicolumn{1}{|c|}{} &
  \multicolumn{1}{c|}{} &
  2 &
  \multicolumn{1}{l|}{\cellcolor[HTML]{E6F3EC}\textbf{3.1}} &
  \multicolumn{1}{l|}{\cellcolor[HTML]{FCFCFF}0} &
  \multicolumn{1}{l|}{\cellcolor[HTML]{FCFCFF}0} &
  \cellcolor[HTML]{FCFCFF}0 \\ \cline{3-7} 
\multicolumn{1}{|c|}{} &
  \multicolumn{1}{c|}{} &
  3 &
  \multicolumn{1}{l|}{\cellcolor[HTML]{B4DFC1}10} &
  \multicolumn{1}{l|}{\cellcolor[HTML]{CAE8D4}6.9} &
  \multicolumn{1}{l|}{\cellcolor[HTML]{E1F1E8}3.8} &
  \cellcolor[HTML]{ACDCBA}\textbf{11} \\ \cline{3-7} 
\multicolumn{1}{|c|}{} &
  \multicolumn{1}{c|}{\multirow{-4}{*}{\begin{tabular}[c]{@{}c@{}}Not\\ Poor\end{tabular}}} &
  4 &
  \multicolumn{1}{l|}{\cellcolor[HTML]{F0F7F5}\textbf{1.7}} &
  \multicolumn{1}{l|}{\cellcolor[HTML]{FCFCFF}0} &
  \multicolumn{1}{l|}{\cellcolor[HTML]{FCFCFF}0} &
  \cellcolor[HTML]{FCFCFF}0 \\ \cline{2-7} 
\multicolumn{1}{|c|}{} &
  \multicolumn{1}{c|}{} &
  1 &
  \multicolumn{1}{l|}{\cellcolor[HTML]{FCFCFF}0} &
  \multicolumn{1}{l|}{\cellcolor[HTML]{FCFCFF}0} &
  \multicolumn{1}{l|}{\cellcolor[HTML]{FCFCFF}0} &
  \cellcolor[HTML]{FCFCFF}0 \\ \cline{3-7} 
\multicolumn{1}{|c|}{} &
  \multicolumn{1}{c|}{} &
  2 &
  \multicolumn{1}{l|}{\cellcolor[HTML]{D9EEE1}\textbf{4.9}} &
  \multicolumn{1}{l|}{\cellcolor[HTML]{FCFCFF}0} &
  \multicolumn{1}{l|}{\cellcolor[HTML]{FCFCFF}0} &
  \cellcolor[HTML]{E7F4ED}3 \\ \cline{3-7} 
\multicolumn{1}{|c|}{} &
  \multicolumn{1}{c|}{} &
  3 &
  \multicolumn{1}{l|}{\cellcolor[HTML]{F2F8F7}1.4} &
  \multicolumn{1}{l|}{\cellcolor[HTML]{EAF5EF}\textbf{2.6}} &
  \multicolumn{1}{l|}{\cellcolor[HTML]{EBF5F0}2.4} &
  \cellcolor[HTML]{FCFCFF}0 \\ \cline{3-7} 
\multicolumn{1}{|c|}{\multirow{-8}{*}{Urban}} &
  \multicolumn{1}{c|}{\multirow{-4}{*}{Poor}} &
  4 &
  \multicolumn{1}{l|}{\cellcolor[HTML]{D6EDDE}\textbf{5.3}} &
  \multicolumn{1}{l|}{\cellcolor[HTML]{FCFCFF}0} &
  \multicolumn{1}{l|}{\cellcolor[HTML]{FCFCFF}0} &
  \cellcolor[HTML]{FCFCFF}0 \\ \hline
\end{tabular}
\end{minipage}
\hfill
\begin{minipage}{0.33\textwidth}
    \caption{\small Percent reduction in average distance traveled (vis-a-vis status quo) by $16$ different groups of people in the portfolio constructed by the portfolio algorithm in Section \ref{sec: monotonically-interpolating-portfolio} to open new pharmacies in Mississippi, USA, with $2\%$ subsidy ($\delta = 0.02$). Each column corresponds to a different $L_p$ norm solution, to open $10$ new facilities to add to existing $206$ CVS, Walgreens, and Walmart pharmacies. Groups are based on rurality, poverty levels, and congressional district. Bolded text represents optimal solutions for different groups (rows). See Section \ref{sec: experiments} for details.}
    \label{tab: spider-plots}
\end{minipage}

\end{table}

\subsection{Outline}

In Section \ref{sec: applications}, we present other applications of our portfolios framework. In Section \ref{sec: related-work}, we discuss related work and previous approaches to portfolios and facility location problems. In Section \ref{sec: portfolios}, we prove Theorems \ref{thm: portfolios-convex-combinations} and \ref{thm: portfolios-lp-norms} that bound portfolio sizes and reduce portfolio problems to approximation algorithm oracles. In Section \ref{sec: approximation-algorithm} we discuss Fair Subsidized Facility Location (FSFL) and prove Theorem \ref{thm: price-of-fairness} giving the approximation algorithm. We also prove Corollary \ref{thm: portfolios-ffl-upper-bound} and Theorem \ref{thm: portfolios-ffl-lower-bound} that give bounds for portfolios for FSFL. Hardness results for FSFL are deferred to Appendix \ref{app: hardness}. Experiments on U.S. Census data are discussed in Section \ref{sec: experiments}, and we conclude in Section \ref{sec: conclusion}. For clarity, we introduce notation and define key concepts as they arise in each section.

\section{Other Applications and Implications of Our Results}\label{sec: applications}

This section showcases several example domains where our portfolio framework applies. The Fair Subsidized Facility Location model is expanded on in Section \ref{sec: approximation-algorithm}; we present other applications here.

\textbf{Bus Routing.} School bus routing is a natural multiobjective optimization problem due to the involvement of several costs (maintaining school buses, fuel costs, etc) and several stakeholders (children, parents, schools). This problem presents different conflicting goals, such as minimizing transportation or fuel costs, the number of buses, and minimizing the wait times for different groups of students (e.g., students belonging to different racial groups) \citep{lee_multi-criteria_1977, bertsimas_optimizing_2019, bertsimas_bus_2020, farhadi_traveling_2021}.

A general approach to dealing with multiple objectives $h_1, \ldots, h_N$ (where $h_1$ is transportation cost, $h_2$ is average student waiting time, $h_3$ is number of buses etc) is to normalize them so their maximum values $\max_{x \in \mathcal{D}} h_i(x) = 1$ for each function $i \in [N]$. Then we define $m > 0$ as the smallest value any $h_i$ takes at any $x \in \mathcal{D}$. {\it Given an algorithm to solve the problem for any given conic combination of $h_1, \ldots, h_N$, Theorem \ref{thm: portfolios-convex-combinations} (Section \ref{sec: convex-combination-portfolio}) then yields $O(1)$-approximate portfolios of size $N \cdot \left(O\left(\log(\frac{N}{m}\right)\right)^{N - 1}$ for this setting for the class $\class = \{\sum_i \lambda_i h_i: \lambda \in \R^N_{\ge 0}\}$ of all conic combinations.}

\textbf{Finance Portfolios.} Consider the following simple model \citep{fabozzi2008portfolio, papahristodoulou_optimal_2004} to trade off risk and rewards in finance portfolio optimization: we are given $n$ assets with mean returns $\mu = (\mu_1, \ldots, \mu_n) \in \R^n$ and variance $\Sigma \in \R^{n \times n}$ for these returns. We seek to diversify our investment among these assets to maximize returns (or minimize regret) while minimizing risk. Formally, we seek a distribution $x \ge 0$ with $\sum_i x_i = 1$ over the assets. One objective is to maximize the return $\mu^\top x$, or equivalently to minimize the reciprocal $h_1(x) = \frac{1}{\mu^\top x}$. The second objective -- that models risk -- is defined as $h_2(x) := \sqrt{x^\top \Sigma x}$.

Let $\sigma_1, \ldots, \sigma_n$ denote the eigenvalues of $\Sigma$, and assume without loss of generality that $\mu_1 \le \mu_2 \le \ldots \le \mu_n$. Then for any feasible $x$, we have $\frac{h_1(x)}{h_2(x)} \le \frac{1}{\mu_1 \min(\sigma_i)} := a$. Similarly, for any feasible $x$, $\frac{h_2(x)}{h_1(x)} \le \mu_n \max_i \sigma_i := b$. Our result (Theorem \ref{thm: portfolios-convex-combinations}) yields an $O(1)$-approximate portfolio of size $O(\log (\max(a, b)))$ for the class $\class = \{\lambda_1 h_1 + \lambda_2 h_2: \lambda_1, \lambda_2 \ge 0\}$ of conic combinations of $h_1, h_2$. Different solutions in the portfolio trade off returns and risk to different degrees.

This is an example of a more general phenomenon in stochastic optimization, where the mean $\E [X]$ (representing returns) of a nonnegative random variable $X$ is often traded off against its second moment $\E [X^2]$ (representing risk or variance). Other examples include healthcare, where treatment effectiveness and risk can be modeled through the first and second moments respectively, and inventory management, where managing expected stockouts (first moment) and maintaining consistent supply (second moment) are often in conflict with each other. More generally, different moments $\E [X^p]$ of $X$ can be traded off against each other. {\it When $X$ has a finite support in $N$ coordinates, our Theorem \ref{thm: portfolios-lp-norms} gives at most $O(\log N)$ solutions while guaranteeing that for all $p \ge 1$, one of these solutions approximates the $p$th moment $\E [X^p]$ within factor $2^p$ of its optimal.} 

\textbf{Fair Representation Clustering.} In this problem \citep{esmaeili_fair_2021, chierichetti_fair_2017}, we are given a metric space $(C, \texttt{dist})$ on clients $C$, an integer $k$, and one of $N$ colors $\chi_j \in [N]$ for each client $j \in C$, and some efficiency constraint. The clients $\{j \in C: \chi_j = t\}$ of color $t \in [N]$ are denoted $\Gamma_t$, and we denote $r_t = \frac{|\Gamma_t|}{|C|}$ to be the fraction of clients of color $t$.

We seek clusters $C(1), \ldots, C(k) \subseteq C$ that partition $C$ while satisfying the efficiency constraint. The goal is to distribute the clients among the clusters in the same ratio as their population: for cluster $C(i)$ and color $t \in [N]$, the deviation $h_t(i)$ for color $t$ in cluster $i$ is defined as
\[
    h_t(i) = \left|r_t - \frac{|C(i) \cap \Gamma_t|}{|C(i)|}\right|,
\]
and we define $h_t = \max_{i \in [k]} h_t(i)$ to be the maximum deviation for color $t$ among all clusters. Given an instance of the Fair Representation Clustering problem, we define $N$ objectives $h_1, \ldots, h_N$ corresponding to the $N$ colors. Previous works have considered the egalitarian/min-max objective $\|\mathbf{h}\|_\infty$ and the utilitarian/min-sum objective $\|\mathbf{h}\|_1$ \citep{esmaeili_fair_2021}. {\it Theorem \ref{thm: portfolios-lp-norms} yields an $O(1)$-approximate portfolio of size $O(\log N)$ for all $L_p$ norm objectives.}

\textbf{Risk-averse Stochastic Programming.} In standard two-stage stochastic programming, we seek to choose a decision vector $x \in \mathcal{D}_x$ in stage 1, after which a \emph{scenario} $\omega \in \Omega$ is realized according to a given distribution $\mathscr{D}$. At this stage, we must choose another decision vector $y \in \mathcal{D}_{y, \omega}$ based on the realized scenario $\omega$. The goal is to minimize $\E_{\mathscr{D}} \left[ G_\omega(x, y) \right]$ for a given convex function $G$ that depends on the realized $\omega$.
% \jme{(Jai to do: add example.)}
% That is, $\E_{\mathscr{D}} \left[ G(x, y) \right] = \int_{\omega} f_{\mathscr{D}}(\omega) G_\omega(x, y)$ where $f_{\mathscr{D}}$ is the probability density function for distribution $\mathscr{D}$.

The distribution $\mathscr{D}$ is often unknown or only partially known ahead of time. In this case, we must make a \emph{risk-averse} \citep{jiang_risk-averse_2018} decision $x$ in the first stage that is effective even for an adversarial distribution $\mathscr{D}$. Since a single solution must hedge against all possible realizations of $\omega$, its worst-case performance is inherently limited. Instead, if we are allowed to choose a \emph{portfolio} of solutions before the scenario $\omega$ is revealed to us, then we can guarantee a good approximation to $G_\omega(x, y)$ irrespective of what distribution $\mathscr{D}$ is. Here, the class of objectives is $\class = \{G_\omega: \omega \in \Omega\}$ indexed by scenarios $\Omega$ and the feasible set is the set of decisions $\mathcal{D}_x$. More generally, using portfolios, we can interpolate between two-stage stochastic and two-stage robust optimization.

\section{Related Work}\label{sec: related-work}

To the best of our knowledge, we are the first to explicitly study portfolios for arbitrary infinite classes of objectives. However, similar notions are implicit in the works of \cite{KK00, goel_simultaneous_2006, GGKT08}. \cite{GGKT08} in particular study $L_p$ norms and their results imply  $O(\log N)$-sized $O(1)$-approximate portfolios for $L_p$ norms. Their technique however crucially uses the structure of $L_p$ norms and does not generalize to other classes that monotonically interpolate between $L_1$ and $L_\infty$ norms.
Our technique for portfolio upper bounds for such families closely resembles the technique of \cite{goel_simultaneous_2006}, who use it to get portfolios for top-$\ell$ norms.

Another well-known related concept is the notion of \emph{Pareto frontier approximations} \citep{papadimitriou_approximability_2000}. Pareto frontier of objectives $h_1, \ldots, h_N$ on feasible set $\mathcal{D}$ is the set of all points $\mathbf{h}(x)$ for $x \in \mathcal{D}$ where not all of the function values can be improved, i.e., for all $x' \in \mathcal{D}, h_i(x') > h_i(x)$ for some $i \in [N]$. The $(1 + \varepsilon)$-approximate Pareto frontier relaxes this constraint by factor $(1 + \varepsilon)$. \cite{papadimitriou_multiobjective_2001} give algorithms to compute Pareto frontiers for several problems, assuming (stronger) bounds $\frac{1}{u} \le h_i(x) \le u$ for all $i \in [N]$ and $x \in \mathcal{D}$, and a stronger feasibility oracle\footnote{They require a polynomial-time oracle to check if the set $\mathcal{D} \cap \{x: h_i(x) \le \lambda_i \ \forall \ i \in [N] \}$ is non-empty for given $\lambda_i, i \in [N]$, while we require only optimizing $\min_{x \in \mathcal{D}}\sum_{i \in [N]} \lambda_i h_i(x)$, which is usually much simpler for many combinatorial problems. For example, if functions $h_i$ represent different path lengths between two vertices in a graph, then the first oracle is NP-hard while the second reduces to another shortest path problem.}. In particular, they obtain $(1 + \varepsilon)$-approximate Pareto frontiers of size $O\left(\left(\frac{\log u}{\varepsilon}\right)^N\right)$, and this is nearly-tight. Note that any Pareto-frontier also implies a portfolio, and therefore, they essentially get a $(1 + \varepsilon)$-approximate portfolio of size $O\left(\left(\frac{\log u}{\varepsilon}\right)^N\right)$. Compared to our result, this is a more restrictive setting, but they obtain a smaller portfolio than what we obtain. 

A long line of works builds on these results to give Pareto frontier approximations for specific settings \citep{bazgan_power_2022, glaser_approximability_2010, herzel_approximation_2021}. In particular, \cite{glaser_approximability_2010} extend the results of \cite{papadimitriou_multiobjective_2001} that use weaker oracles for conic combinations, similar to our result, while still considering the scaling assumption on individual function values. They construct a Pareto frontier (and therefore a portfolio) of size $O\left(\left(\frac{\log u}{\varepsilon}\right)^N\right)$, and one can show this is an $N(1 + \varepsilon)$-approximate. Note that their result implies that every factor decrease in approximation quality $\epsilon^\prime = \epsilon/\gamma$ (for $\gamma > 1$) increases the size of the portfolio exponentially by $\gamma^N$. Our setting on the other hand uses a weaker assumption\footnote{For any function that satisfies $1/u \leq h_i(x) \leq u$, it also satisfies $h_i(x)/h_j(x) \leq u^2$ for all $x\in \mathcal{D}$.} that $\frac{h_i(x)}{h_j(x)} \le u$ for all $x\in D$, for all $i, j \in [N]$ and generalizes their setting. Further, our Theorem \ref{thm: portfolios-convex-combinations} improves upon the approximation ratio of \cite{glaser_approximability_2010} {by a factor of $N$, producing a $(1 + \varepsilon)$-approximate portfolio, without increasing the size exponentially by around $N^N$. Our portfolio is only of a slightly} larger size $N\cdot\left(\frac{O\left(\log(Nu/\varepsilon)\right)}{\varepsilon}\right)^{N - 1}$. The main technical difference in our approach is that we use a combination of a multiplicative and additive $\varepsilon$-meshes, as opposed to the above results that use only a multiplicative mesh.

The notions of Pareto frontiers and portfolios coincide in the special case of size equal to $1$, i.e, if some feasible solution $x^* \in \mathcal{D}$ is simultaneously optimal for all $h_i(\cdot), i \in [N]$, then the Pareto frontier reduces to the single point $\{\mathbf{h}(x^*)\}$. This also holds true when $x^* \in \mathcal{D}$ is \emph{simultaneouly $\alpha$-approximate} for all $h_i(\cdot)$ or that $h_i(x^*) \le \min_{x \in \mathcal{D}} h_i(x)$ for all $i$, in which case the single point $\{\mathbf{h}(x^*)\}$ coordinate-wise $\alpha$-approximates the Pareto frontier. Simultaneous approximations have been extensively studied (e.g. \cite{KK00, AERW04, goel_simultaneous_2006}). \cite{goel_simultaneous_2006} in particular study simultaneous approximations for various problems and establish a fundamental result that shows that simultaneous $\alpha$-approximations for top-$\ell$ norms are simultaneous $\alpha$-approximations for other classes of norms like $L_p$ norms. {When portfolio size is greater than $1$, portfolios for top-$\ell$ norms may not be portfolios for $L_p$ norms (see Appendix \ref{app: portfolios-are-not-transferable} for an example).}

$L_p$ norm objectives are widely considered in the approximation algorithms literature as a model for fairness and as interesting theoretical questions \citep{AYZ04, GGKT08, farhadi_traveling_2021, MSV21, CMV21, KK00}. In particular, \cite{GGKT08} study fixed norm objectives including $L_p$ norm objectives for $k$-Clustering. Our FSFL model thus generalizes their setting using the subsidy constraint (see Appendix \ref{app: clustering-and-ufl}). Variants of both Uncapacitated Facility Location \citep{hochbaum1982heuristics, CNW83, STA97, sviridenko2002improved, GK98, KGPR98, CG99, KMS02, JMMSV03, CFS03, Byrka07, Li13} and $k$-Clustering \citep{LV92, arora1998approximation, charikar1999constant, SO13, cohen2022improved, chakrabarty2022approximation} are very well-studied in the Operations Research and Computer Science literature.
While all of these algorithms appropriately bound the number/cost of open facilities, they do not bound the total loss of facilities, which is where our new rounding subroutine is needed. Many other kinds of fairness criteria in facility location problems are well-studied from both theoretical and applied perspectives \citep{Truelove93, Talen98, jung2019center, ABV21, MSV21} etc.
Most of these models either fix a fairness criterion or do not discuss how a choice among a suite of fairness criteria must be made.

\section{Portfolios}\label{sec: portfolios}

In this section, we consider the following setting: we are given $N$ positive functions $h_1, \ldots, h_N: \mathcal{D} \to \mathbb{R}_{> 0}$ over some domain $\mathcal{D}$ of feasible points that represent different objectives we seek to minimize. We refer to these as the \emph{base objective functions}. These objectives are often competing, i.e., no single point $x \in \mathcal{D}$ minimizes each $h_i, i \in [N]$ simultaneously, even approximately. We will denote $\mathbf{h}(x) = (h_1(x), \ldots, h_N(x)) \in \R_{> 0}^N$ for all $x \in \mathcal{D}$. First, we construct portfolios for the class of all conic combinations of $h_1, \ldots, h_N$ in Section \ref{sec: convex-combination-portfolio}, and then in Section \ref{sec: monotonically-interpolating-portfolio} discuss portfolios for classes of functions that monotonically interpolate between the $L_1$ norm objective $\|\mathbf{h}\|_1 := h_1 + \ldots + h_N$ and the $L_\infty$ norm objective $\|\mathbf{h}\|_\infty := \max_{i \in [N]} h_i$.

\subsection{Portfolios for Conic Combinations}\label{sec: convex-combination-portfolio}

Our main result gives a portfolio in terms of the \emph{imbalance} of the base objective functions, defined as follows: we say that $N$ positive functions $h_1, \ldots, h_N: \mathcal{D} \to \mathbb{R}_{> 0}$ are $u$-balanced for some $u \ge 1$ if for all $x \in \mathcal{D}$ and for all $i, j \in [N]$, $\frac{h_i(x)}{h_j(x)} \le u$.

\convexCombinationsPortfolio*

We present the proof for the case $\beta = 1$, i.e., when the oracle returns the optimal solution for each $g_\lambda := \sum_{j} \lambda_j h_j$. The generalization to the case $\beta > 1$ is straightforward and left to the reader.

\begin{figure}
    \centering
    \includegraphics[width=0.5\linewidth]{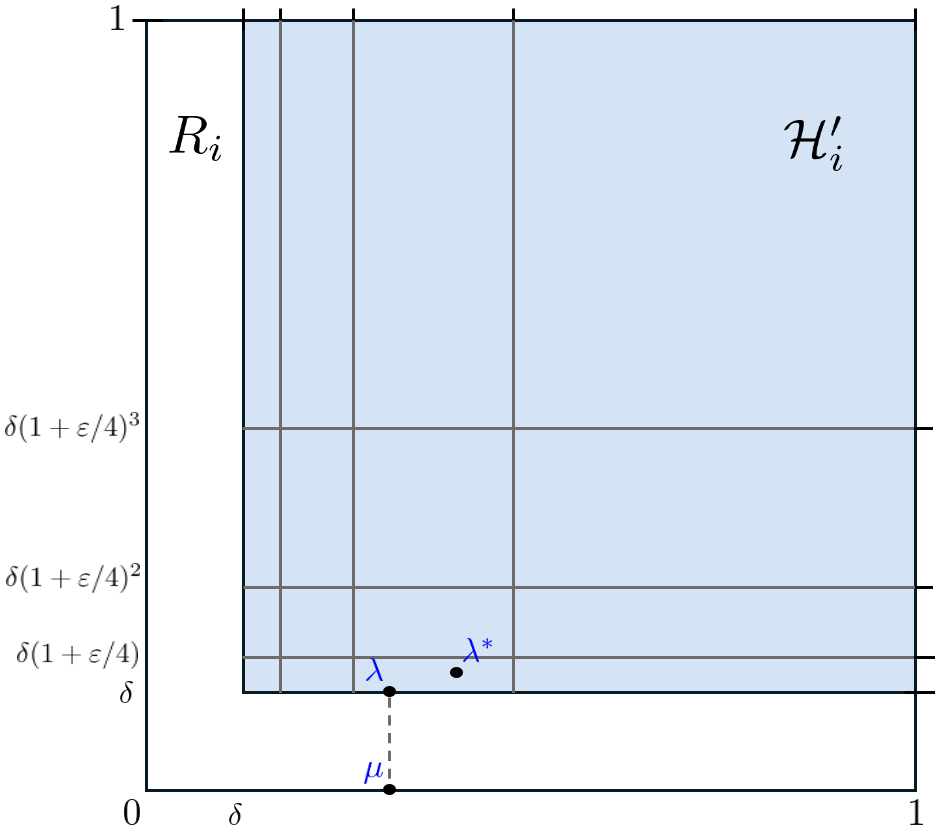}
    \caption{An illustration for the mesh for $\mathcal{H}_i$ used in the proof of Theorem \ref{thm: portfolios-convex-combinations}.}
    \label{fig: mesh}
\end{figure}
 
\emph{Proof.} The plan is as follows: for each $i \in [N]$, define the $(N - 1)$-dimensional hypercube $\mathcal{H}_i := \{\lambda \in [0, 1]^N: \lambda_i = 1\}$. We show first that it is sufficient to restrict to convex combinations $g_\lambda := \sum_{i \in [N]} \lambda_i h_i$ where $\lambda \in \mathcal{H}_1 \cup \mathcal{H}_2 \cup \ldots \cup \mathcal{H}_N$ by rescaling the highest coordinate of $\lambda$ to be $1$. We will partition each $\mathcal{H}_i$ using an $\varepsilon$-parameterized \emph{mesh} and choose a single representative $\lambda^*$ for each part of the mesh so that the optimal point $x(\lambda^*) = {\arg\min}_{x \in \mathcal{D}} g_{\lambda^*}(x)$ for $g_{\lambda^*} := \sum_{j \in [N]} \lambda^*_j h_j$ will be a $(1 + \varepsilon/4)$-approximate solution for $g_{\lambda}$ for every other $\lambda$ in the part. Bounding the number of these parts will give the desired bound on the portfolio size.

To create this mesh, we will find some $\delta > 0$ and partition $\mathcal{H}_i = \mathcal{H}_{i}' \cup R_{i}$ where $\mathcal{H}_{i}' = \{\lambda \in [\delta, 1]^N: \lambda_i = 1\}$ is a smaller $(N - 1)$-dimensional hypercube contained within $\mathcal{H}_i$, and $R_i = \mathcal{H}_{i} \setminus \mathcal{H}_{i}'$ are the remaining points in $\mathcal{H}_i$. On this smaller hypercube $\mathcal{H}'_i$, we put the standard \emph{multiplicative} mesh that partitions it into hyper-rectangles (see Figure \ref{fig: mesh}). That is, we define nested boxes using \emph{thresholds} $\{\delta, \delta(1 + \varepsilon/4), \delta(1 + \varepsilon/4)^2, \ldots, 1\}$ for each coordinate (not equal to $i$). The number of thresholds is $T + 1 = \log_{1 + \varepsilon/4}(1/\delta) + 1$ for each dimension, and for every $\lambda \in \mathcal{H}_i'$, we can assign it to one of the $T^{N-1}$ boxes based on where the coordinates of $\lambda$ lie within these thresholds, i.e., $\lambda \in  [\delta(1+\varepsilon/4)^{k_1}, \delta(1+\varepsilon/4)^{k_1+1}] \times \ldots \times [\delta(1+\varepsilon/4)^{k_N}, \delta(1+\varepsilon/4)^{k_N+1}]$, for appropriate $k_1, \ldots, k_N$.

For the remaining points $R_i = \mathcal{H}_i \setminus \mathcal{H}_i'$, we create an \emph{additive} mesh. The key observation is the following: for all $\lambda, \mu \in \R^N_{\ge 0}$, the optimal point $x(\lambda) := {\arg\min}_{x \in \mathcal{D}} g_\lambda(x)$ for the linear combination $g_\lambda = \sum_{j} \lambda_j h_j$ is an approximate solution for $g_\mu$, and the quality of this approximation can be bounded in terms of the $L_1$ norm distance $\|\lambda - \mu\|_1$. Each $\mu \in R_i$ can first be mapped to some $\lambda \in \mathcal{H}_i'$ with $\|\lambda - \mu\|_1 \le 2N\delta$, and then this $\lambda$ can be mapped to the corresponding $\lambda^*$ in the multiplicative mesh. As we show later, for the appropriate choice of $\delta$, the first step loses approximation factor $(1 + \varepsilon/4)$. The second step loses factor $(1 + \varepsilon/4)^2$, and thus the final approximation factor is at most $(1 + \varepsilon/4)^3 \le 1 + \varepsilon$ for $\varepsilon \in (0, 1]$. We choose $\delta = \frac{\varepsilon}{4uN}$, and we show the remaining claims next. 

\textbf{Approximation guarantee.} First, say $\lambda \in \mathcal{H}'_i$. Then, by construction of the multiplicative mesh, we have that any $\lambda^*$ in the same part of the mesh satisfies that $\frac{1}{1 + \varepsilon/4} \le \frac{\lambda_j}{\lambda^*_j} \le 1 + \varepsilon/4$. This implies that $x(\lambda^*)$ is a $(1 + \varepsilon/4)^2$-approximation for $g_\lambda$ using the optimality of $x(\lambda^*)$ for $g_{\lambda^*}$:
\begin{align}
    g_\lambda(x(\lambda^*)) &= \sum_j \lambda_j h_j(x(\lambda^*)) \le (1 + \varepsilon/4) \sum_j \lambda^*_j h_j(x(\lambda^*)) \le (1 + \varepsilon/4) \sum_j \lambda^*_j h_j(x(\lambda)) \label{eqn: multiplicative-mesh-inequality} \\
    &\le (1 + \varepsilon/4)^2 \sum_j \lambda_j h_j(x(\lambda)) = (1 + \varepsilon/4)^2 \min_{x \in \mathcal{D}} g_\lambda(x). \notag
\end{align}

Now, suppose we pick some $\mu \in R_i$. Define $\lambda = (\lambda_1, \ldots, \lambda_N) = (\max(\mu_1, \delta), \ldots, \max(\mu_N, \delta))$. Then $\lambda \in \mathcal{H}'_i$. We will show using a sequence of inequalities that $x(\lambda^*)$ is a $(1 + \varepsilon/4)^3$-approximation for $g_\mu$, where as above $\lambda^*$ is the representative point of the part of the mesh that contains $\lambda$. To show this, we need the following claim:

\begin{claim}\label{claim: additive-mesh}
    For all $i, j \in [N]$, all $\lambda \in \mathcal{H}_i$ and all $x \in \mathcal{D}$, we have $h_j(x) \le u \cdot g_\lambda(x)$.
\end{claim}

\emph{Proof.}
    Note that $g_\lambda(x) = \sum_{\ell \in [N]} \lambda_{\ell} h_{\ell}(x) = \|\lambda\|_1 \sum_{\ell \in [N]} \frac{\lambda_{\ell}}{\|\lambda\|_1} h_{\ell}(x)$. However, $\sum_\ell \frac{\lambda_{\ell}}{\|\lambda\|_1} h_{\ell}(x)$ is a convex combination of the base objectives $h_1(x), \ldots, h_N(x)$. Since the base objectives are $u$-balanced, $\sum_\ell \frac{\lambda_{\ell}}{\|\lambda\|_1} h_{\ell}(x) \ge \sum_\ell \frac{\lambda_{\ell}}{\|\lambda\|_1} \left(\frac{1}{u} \cdot h_j(x)\right) = \frac{h_j(x)}{u}$. This implies that $g_\lambda(x) \ge \|\lambda\|_1 \frac{h_j(x)}{u}$, or that $h_j(x) \le \frac{u \cdot g_\lambda(x)}{\|\lambda\|_1}$. Since $\lambda \in \mathcal{H}_i$, the $i$th coordinate $\lambda_i = 1$, so that $\|\lambda\|_1 \ge 1$. \qed

With this claim, we have the following bound on the performance of $x(\lambda^*)$ for $g_\mu$:
\begin{align*}
    g_\mu(x(\lambda^*)) &= \sum_{j \in [N]} \mu_j h_j(x(\lambda^*)) = \sum_{j \in [N]} (\lambda_j + (\underbrace{\mu_j - \lambda_j}_{\le 0})) h_j(x(\lambda^*)) \\
    &\le \sum_{j \in [N]} \lambda_j h_j(x(\lambda^*)) = g_\lambda(x(\lambda^*)).
\end{align*}
However, from eqn. (\ref{eqn: multiplicative-mesh-inequality}), $g_\lambda(x(\lambda^*)) \le (1 + \varepsilon/4)^2 g_\lambda(x(\lambda))$. Using optimality of $x(\lambda)$ for $g_\lambda$,
\begin{align*}
    g_\lambda(x(\lambda)) &\le g_\lambda(x(\mu)) = \sum_{j \in [N]} (\mu_j + (\lambda_j - \mu_j)) h_j(x(\mu)) = g_\mu(x(\mu)) + \sum_{j}  (\lambda_j - \mu_j) h_j(x(\mu))
\end{align*}
Finally, we use Claim \ref{claim: additive-mesh} bounding $h_j(x(\mu)) \le u \cdot g_\mu(x(\mu))$, so that
\begin{align*}
    g_\lambda(x(\lambda)) - g_\mu(x(\mu)) &\le \sum_{j} (\lambda_j - \mu_j) \times u \cdot g_\mu(x(\mu)) = u \cdot g_\mu(x(\mu)) \sum_{j} (\lambda_j - \mu_j) \le u \cdot g_\mu(x(\mu)) \cdot N\delta
\end{align*}
since $0 \le \lambda_j - \mu_j \le \delta$ for all $j$. Since $\delta = \frac{\varepsilon}{4uN}$, we get that $g_\lambda(x(\lambda)) \le (1 + \varepsilon/4) g_\mu(x(\mu))$. Together,
\[
    g_\mu(x(\lambda^*)) \le g_\lambda(x(\lambda^*)) \le (1 + \varepsilon/4)^2 g_\lambda(x(\lambda)) \le (1 + \varepsilon/4)^3 g_\mu(x(\mu)) \le (1 + \varepsilon) g_\mu(x(\mu)).
\]

\textbf{Portfolio size guarantee.} Each $\mathcal{H}_i$ is an $(N - 1)$-dimensional hypercube, and so its multiplicative mesh has size $\left(\log_{1 + \varepsilon/4}(1/\delta)\right)^{N - 1} \le \left(\frac{12}{\varepsilon} \log(1/\delta)\right)^{N - 1} = \left(\frac{12}{\varepsilon} \log\left(\frac{4uN}{\varepsilon}\right)\right)^{N - 1}$. The portfolio size is upper bounded by the union of the sizes of the meshes:
\[
    N \times \left(\log_{1 + \varepsilon/4}(1/\delta)\right)^{N - 1} \le N \left(\frac{12}{\varepsilon} \log\left(\frac{4uN}{\varepsilon}\right)\right)^{N - 1}. \quad \textup{\qed}
\]

\subsection{Portfolios for Interpolating Functions}\label{sec: monotonically-interpolating-portfolio}

In this section, given base objective functions $h_1, \ldots, h_N: \mathcal{D} \to \R_{\ge 0}$ over a domain $\mathcal{D}$ of feasible solutions, we prove Theorem \ref{thm: portfolios-lp-norms} that obtains portfolios for classes $\class$ of functions that interpolate monotonically between the $L_1$ norm objective $\|\mathbf{h}\|_1 := h_1 + \ldots + h_N$ and the $L_\infty$ norm objective $\|\mathbf{h}\|_\infty := h_1 + \ldots + h_N$.

Formally, given reals $a \le b$, the class $\class = \{g_\lambda: \lambda \in [a, b]\}$ of objectives on $\mathcal{D}$ interpolates monotonically between $\|\mathbf{h}\|_1$ and $\|\mathbf{h}\|_\infty$ if (1) $g_a = \|\mathbf{h}\|_1$, (2) $g_b = \|\mathbf{h}\|_\infty$, and (3) $g_\lambda$ is non-increasing with $\lambda$, i.e., for all $\mu \ge \lambda$ and $x \in \mathcal{D}$, we must have $g_\lambda(x) \ge g_\mu(x)$. Examples of such a class include $L_p$ norm functions $\left\{\|\mathbf{h}\|_p := \left(\sum_{i \in [N]} h_i^p\right)^{1/p}: p \ge 1\right\}$, convex combinations of $\|\mathbf{h}\|_1$ and $\|\mathbf{h}\|_\infty$, and top-$\ell$ norm functions for $\ell \in [N]$ that sum the $\ell$ highest coordinates. 

We show that given an oracle to find a $\beta$-approximate solution for any given $h_\lambda \in \class$, we can obtain a $\beta(1 + \varepsilon)$-approximate portfolio of size $O\left(\frac{\log \beta N}{\varepsilon}\right)$ in a polynomial number of oracle calls.

\monotonicInterpolationPortfolio*

Given this result, one might naturally ask if portfolios for one class of monotonically interpolating norms (e.g., top-$\ell$ norms) are also portfolios for another such class (e.g., $L_p$ norms); indeed, a portfolio of size $1$ that is $\alpha$-approximate for top-$\ell$ norms is $\alpha$-approximate for $L_p$ norms \citep{goel_simultaneous_2006}.
In Appendix \ref{app: portfolios-are-not-transferable}, we show that this is false for portfolios of size $> 1$. That is, there exist portfolios for top-$\ell$ norms which are not approximate portfolios for $L_p$ norms.

\emph{Proof of Theorem \ref{thm: portfolios-lp-norms}.} Denote the class $\mathbf{C} = \{g_\lambda: \lambda \in [a, b]\}$. We assume we are given an oracle to obtain $\beta$-approximation for $g_\lambda$ over $\mathcal{D}$ for any given $\lambda \in [a, b]$. In a polynomial number of oracle calls, we will find a $\beta(1 + \varepsilon)$-approximate portfolio for $\mathbf{C}$. The portfolios size is $S + 1$ where $S = \log_{1 + \varepsilon} (\beta N)$.

We construct a sequence of $\lambda$ values $a = \lambda(0) < \lambda(1) < \ldots < \lambda(S) = b$ such that the set $P$ of $\beta$-approximate solutions for the $S + 1$ objective functions $g_{\lambda(0)}, g_{\lambda(1)}, \ldots, g_{\lambda(S)}$ forms the desired portfolio. For $\lambda \in [a, b]$, let $\OPT_\lambda := \min_{x \in \mathcal{D}} g_\lambda(x)$ be the optimal value for ${g_\lambda}$, with $x(\lambda)$ denoting the $\beta$-approximate solution returned by the oracle. The objective value of this solution for $g_\lambda$ is denoted $\ALG_\lambda$.
By definition, $\ALG_{\lambda} \le \beta \cdot \OPT_{\lambda}$. 

We can assume without loss of generality that $\ALG_\lambda$ is non-increasing with $\lambda$: indeed, given $\mu > \lambda$, the cost of $x(\lambda)$ for $g_\mu$ is $g_\mu(x(\lambda)) \le g_\lambda(x(\lambda)) = \ALG_\lambda$ by the monotonicity assumption. Therefore, if $g_\mu(x(\mu)) > g_\lambda(x(\lambda))$, we can use $x(\lambda)$ instead of $x(\mu)$ with better a objective value for $g_\mu$.

Next, for $\lambda \in [a, b]$, denote by $\lambda'$ the minimum value of $\mu \in [a, b]$ such that $\frac{\ALG_{\lambda}}{(1 + \varepsilon)} \ge \ALG_{\mu}$. Intuitively, we take a `step' of size $(1 + \varepsilon)$ in the objective value.
If no such $\mu$ exists (i.e.,  when ${\ALG_{b}} > \frac{\ALG_{\lambda}}{1 + \varepsilon}$), define $\lambda' = b$. Construct portfolio $P_\varepsilon$ as follows: initially set $\lambda(0) = a$ and $i = 0$. While $\lambda(i)' < b$, keep taking $(1 + \varepsilon)$-steps, i.e., setting $\lambda(i + 1) = \lambda(i)'$ and increasing the counter $i$. Suppose $\lambda(0), \ldots, \lambda(S)$ is the sequence of $\lambda$ values generated by this algorithm. The algorithm outputs the corresponding $\beta$-approximations, i.e., $P_\varepsilon = \left\{x(\lambda(i)): i \in [0, S]\right\}$.

We claim that for each $i \in [0, S - 1]$ and $\mu \in [\lambda(i), \lambda(i + 1))$, solution $x(\lambda(i))$ is a $\beta (1 + \varepsilon)$-approximation to $g_{\mu}$. This is sufficient to prove the approximation guarantee of the portfolio. Since $g_\mu \le g_{\lambda(i)}$, the cost of $x(\lambda(i))$ for ${g_{\mu}}$ is $g_\mu(x(\lambda(i)) \le g_{\lambda(i)}(x(\lambda(i)) = \ALG_{\lambda(i)}$.
Now, by definition of $\lambda(i + 1)$, we have $\ALG_{\mu} \ge \frac{\ALG_{\lambda(i)}}{1 + \varepsilon}$, so that this cost is at most $(1 + \varepsilon) \ALG_{\mu} \le (1 + \varepsilon) \beta \cdot \OPT_\mu$. This proves the approximation guarantee for $P_\varepsilon$.

We now prove that $|P| = O(\log_{1 + \varepsilon} (\beta N))$. By construction, $\ALG_{\lambda(i + 1)} \le \ALG_{\lambda(i)}/(1 + \varepsilon)$, i.e., the value of $\ALG_{\lambda(i)}$ decreases by factor $\ge (1 + \varepsilon)$ in each step (except possibly the last). Therefore, the number of steps $S$ is bounded by $\log_{(1 + \varepsilon)} \frac{\ALG_{a}}{\ALG_b}$. It is now sufficient to prove that $\ALG_a \le N\beta \cdot \ALG_b$. To see this claim, since $x(a)$ is a $\beta$-approximation to $g_a = \|\mathbf{h}\|_1$, we have $\ALG_a = \|\mathbf{h}\|_1(x(a)) \le \beta \cdot \|\mathbf{h}\|_1(x(b))$. Next, since $\|\mathbf{h}\|_1(x(b)) = h_1(x(b)) + \ldots + h_N(x(b)) \le N \cdot \|\mathbf{h}\|_\infty(x(b))$, we get $\ALG_a \le \beta N \cdot \ALG_b$ since $g_b$ is the $L_\infty$ norm. \qed

\section{Fair Subsidized Facility Location}\label{sec: approximation-algorithm}

In this section, we construct a polynomial-time approximate oracle (and consequently portfolios) for the Fair Subsidized Facility Location (FSFL) problem, where the input consists of (1) a metric space $(X, \texttt{dist})$ on set $X = C \cup F$ of clients $C$ and potential facilities $F$, (2) nonnegative operating costs $c: F \to \R_{\ge 0}$ for facilities, (3) nonnegative revenues $r: C \to \R_{\ge 0}$ for clients, (4) \emph{subsidy} parameter $\delta > 0$, and (5) an objective function $g: \R^C \to \R$ on client distances that is convex and sublinear, i.e., for all $d, d' \in \R^C$ and $\alpha \ge 0$, we must have $g(\alpha d + d') \le \alpha g(d) + g(d')$. A feasible solution is a pair $(F', \Pi)$ where (1) $F' \subseteq F$ is a set of open facilities and (2) $\Pi: C \to F'$ are client assignments to open facilities.

The mild conditions of convexity and sublinearity on the objective lead to a rich class of objectives: it includes classical objectives such as the sum of client distances and the maximum client distance, as well as norms of client group distances when client groups are specified. In particular, this includes the $L_p$ norms of client group distances, where we are given $t$ groups through fractional group memberships $\mu_{j, s} \ge 0$ for each client $j \in C$ and each group $s \in [t]$. For solution $(F', \Pi)$, the group distance of the $s$th group is $h_s^{(\Pi)} := \sum_{j \in C} \mu_{j, s} \texttt{dist}_{j, \Pi(j)}$ and the $L_p$ norm objective for given $p \ge 1$ is to minimize $\|\mathbf{h}^{(\Pi)}\|_p := \left(\sum_{s \in [t]}(h^{(\Pi)}_s)^p\right)^{1/p}$. Formally, our result states the following:

\approximationAlgorithm*

\begin{table}[]
\centering
\caption{A summary of various steps in the rounding algorithm for FSFL.}
\label{tab: rounding-algorithm}
\setlength\fboxsep{1pt} % Optional: adjust padding between box and content
\fbox{%
\hspace*{-3.5cm}%
\begin{tabular}{c|c|c|c|c}
    Solution                       & $x$        & $y$        & \begin{tabular}[c]{@{}c@{}}Distance\\ approximation\end{tabular} & Subsidy             \\
    $(x, y)$                       & fractional & fractional & $1$                                                              & $\delta$            \\
    $\phantom{\small \alpha\textsc{-PointRounding}}\bigg\downarrow \small \alpha\textsc{-PointRounding}$ &
       &
       & \small (Lemma \ref{lem: alpha-point-rounding}) 
      & \small (Lemma \ref{lem: alpha-point-rounding}) \\
    $(\overline{x}, \overline{y})$ & fractional & fractional & $4 \cdot \max\left(1, \frac{1}{\delta}\right)$                                      & $2\delta$           \\
    $\small \phantom{\textsc{RoundToIntegralFacilities}}\bigg\downarrow \small\textsc{RoundToIntegralFacilities}$ &
       &
       &
      \small (Lemma \ref{lem: integral-facilities-bounded-distances}) &
      \small (Lemma \ref{lem: integral-facility-bounded-loss}) \\
    $(x', y')$                     & fractional & integral   & $20 \cdot \max\left(1, \frac{1}{\delta}\right)$                                     & $2\delta$           \\
    $\small \phantom{\textsc{IntegralAssignment}}\bigg\downarrow \small\textsc{IntegralAssignment}$ &
       &
       &
      \small (Lemma \ref{lem: integral-assignment}) &
      \small (Lemma \ref{lem: integral-assignment}) \\
    $(x'', y')$                    & integral   & integral   & $20 \cdot \max\left(1, \frac{1}{\delta}\right)$                                     & $2 \delta + \theta$ \\
\end{tabular}%
}
\end{table}

We start by writing a convex relaxation for the problem. We represent solutions to the above problem through characteristic vectors: the set $F' \subseteq F$ of open facilities is represented by the binary vector $y \in \{0, 1\}^F$ such that $y_f = 1$ if and only if $f \in F'$. Similarly, an assignment $\Pi: C \to F'$ is represented by the vector $x \in \R^{C \times F}$ such that $x_{j, f} = 1$ if and only if $\Pi(j) = f$ for all clients $j \in C$ and $f \in F$.

These vectors satisfy two natural constraints: (1) a client is only assigned to an open facility, i.e., $x_{j, f} \le y_f$ for all $j \in C, f \in F$ and (2) each client $j \in C$ must be assigned to some facility, or $\sum_{f \in F} x_{j, f} = 1$. This is useful for writing convex relaxations of the problem, where we instead allow $x_{j, f}$ and $y_f$ to be in $[0, 1]$ for each client $j \in C$ and facility $f \in F$ under the above constraints \citep{LV92, STA97}. Further, the distance $d_j$ of a client $j \in C$ to its assigned facility can be written as $\sum_{f \in F} x_{j, f} \texttt{dist}_{j, f}$.

We introduce a variable $\ell_f$ for each $f \in F$ to denote the loss of the facility $f$, and impose the constraint $\ell_f \ge \max\left(0, y_f c_f - \sum_{j \in C} x_{j, f} r_j\right)$. That is, the $\ell_f = 0$ when $y_f = 0$ or when $f$ is not open. When $f$ is open but profitable, i.e., revenue $\sum_{j \in C} x_{j, f} r_j$ of clients assigned to it exceeds the operating cost $c_f y_f$, then loss $\ell_f = 0$, otherwise $\ell_f$ is the difference in the operating cost and revenue. The final constraint is the $\delta$-subsidy constraint that bounds the total loss $\sum_{f \in F} \ell_f$ to be at most $\delta \sum_{j \in C} r_j$. 
\begin{align}
    \min &\ g(\dist) := \left(\sum_{s \in [t]}\left(\sum_{j\in C} \mu_{j,s} d_j\right)^p \right)^{1/p}\label{eqn: price-of-fairness-ip}\tag{IP} \\
    \text{s.t. } &\dist_j = \sum_{f \in F} x_{j, f} \texttt{dist}_{j, f}, &\forall \ j \in C, \\
    &\sum_{f \in F} x_{j, f} = 1, &\forall \ j \in C, \label{cons: assign-every-client}  \\
    &x_{j, f} \le y_f, &\forall \ j \in C, f \in F, \label{cons: assign-clients-only-to-open-facilities} \\
    &\loss_f \ge c_f y_f - \sum_{j \in C} x_{j, f} r_j, &\forall \ f \in F, \label{eqn: loss-is-more-than-cost-gap} \\
    &\sum_{f \in F} \loss_f \le \delta \sum_{j \in C} r_j, \label{eqn: subsidy-constraint} \\
    &\loss_f \ge 0, &\forall \ f \in F, \label{eqn: loss-is-nonnegative} \\
    &x \in \{0, 1\}^{C \times F}, y \in \{0, 1\}^F. \notag
\end{align}
Though we specify the objective function $g(\cdot)$ to be some $L_p$ norm of the group distances, our analysis only requires $g(\cdot)$ to be convex and sublinear. Due to the convexity of $g(\cdot)$, we can relax the integrality constraints on $x, y$ to get a convex program and obtain an optimal \emph{fractional solution} $(x, y, \loss, \dist)$ in polynomial time. Since $\loss$ and $\dist$ can be determined using $x, y$, we will often omit them and denote the solution as $(x, y)$. When both $x, y$ are integral, we call $(x, y)$ an \emph{integral} solution. In this case, $(x, y)$ corresponds to a solution $(F', \Pi)$ of the original problem. Since $g$ is sublinear, we have that $g(\gamma\dist) \le \gamma g(\dist)$ for all $\gamma > 0$, and since $\dist$ is linear in $x$, a rounding algorithm that increases $x$ by factor $\gamma$ increases the objective $g$ by factor $\le \gamma$.

Given the optimal fractional solution $(x, y)$ to the convex program, in Section \ref{sec: bounded-distances} we round it to a solution $(x', y')$ where $y'$ is integral (but $x'$ may be fractional) using algorithms $\alpha$-\textsc{PointRounding} and \textsc{RoundToIntegralFacilities} respectively. Then, we round $(x', y')$ to an integral solution $(x'', y')$ in Section \ref{sec: rounding-last-step}, using a subroutine for the generalized assignment problem from \cite{shmoys_approximation_1993}. Table \ref{tab: rounding-algorithm} illustrates these steps in the rounding algorithm. Section \ref{sec: ffl-portfolios} discusses portfolios for FSFL.

\subsection{Finding an Integral Set of Facilities}\label{sec: bounded-distances}

The first step in our rounding procedure is the $\alpha$-\textsc{PointRounding} subroutine from \cite{STA97}. For appropriately chosen $\alpha \in (0,1)$, and fractional optimal $(x,y)$, this algorithm removes the assignment of clients $j \in C$ to facilities that are the farthest and hold at most $(1-\alpha)$ fraction of the assignment for $j$. The algorithm then rescales the $y$ variables by $1/\alpha$ while maintaining feasibility to the convex relaxation of (\ref{eqn: price-of-fairness-ip}). We state the algorithm in Appendix \ref{sec: alpha-point-rounding} for completeness. Here, we state the formal guarantees of $\alpha$-\textsc{PointRounding} for facility location. Given a vector $\Delta \in \R_{\ge 0}^C$, we say that a fractional solution $(\hat{x}, \hat{y})$ is $\Delta$-close if for all clients $j \in C$ and all facilities $f \in F$ such that $\hat{x}_{j, f} > 0$, we have $\texttt{dist}_{j, f} \le \Delta_j$ (note that $\Delta_j$ need not all be the same value for all $j$). Note that for any fractional solution $(x,y)$, the actual distance $\texttt{\textup{dist}}_{j,f}$ of a client $j$ to any facility $f$ that they are fractionally matched to could be much larger than the expected distance $d_j = \sum_{f' \in F}\texttt{dist}_{j, f'} x_{j, f'}$ that the convex program pays for. We show in the following lemma that this can be bounded, without violating the subsidy constraint by too much:

\begin{restatable}{lemma}{alphaPointRounding}\label{lem: alpha-point-rounding}
    Fractional solution $(\overline{x}, \overline{y})$ output by \textsc{\textup{$\alpha$-PointRounding}$(x, y)$} (Algorithm \ref{alg: alpha-point-rounding}) satisfies: 
    \begin{enumerate}
        \item It is $\Delta$-close where $\Delta_j \le 4 \max\left(1, \frac{1}{\delta}\right) \dist_j$ for all $j \in C$, where $d_j = \sum_f x_{j,f} \texttt{\textup{dist}}_{j,f}$ is the expected distance in $(x,y)$. That is, for any facility $f$ with $\overline{x}_{j, f} > 0$,
        \[
            \overline{x}_{j, f} >0 \implies \texttt{\textup{dist}}_{j, f} \le 4 \max\left(1, \frac{1}{\delta}\right) \dist_j.
        \]
        \item The total loss of $(\overline{x}, \overline{y})$ is at most fraction $2\delta$ of the total revenue, i.e., $\sum_{f \in F} \overline{\loss_f} \le 2 \delta \sum_{j \in C} r_j$.
    \end{enumerate}
\end{restatable}

Crucially, Part 1 states that a client $j$ can now be assigned to \emph{any} facility $f$ such that $\overline{x}_{j, f} > 0$ while being within factor $4 \max\left(1, \frac{1}{\delta}\right)$ of its original distance $d_j$ in the fractional solution $(x, y)$. This part of the lemma follows similar arguments to \cite{STA97}. Part 2 of the lemma requires tracking the losses from the original solution $(x, y)$ to the output solution $(\overline{x}, \overline{y})$; we include details of the proof in Appendix \ref{sec: omitted-proofs-algorithm}. Hereafter, we denote this set of `feasible' facilities for client $j \in C$ as
\[
    \feas_j := \{f \in F: \overline{x}_{j, f} > 0\}.
\]
Lemma \ref{lem: alpha-point-rounding} implies that if we open a facility in $\feas_j$ for each client $j \in C$ and assign $j$ to this facility, then we obtain a solution where the objective $g(\cdot)$ is within a factor $4\max\left(1, \frac{1}{\delta}\right)$ of the relaxed optimal of (\ref{eqn: price-of-fairness-ip}). However, opening a (potentially) different facility for each client can significantly increase the total operating cost of open facilities, and therefore the total loss. Therefore, we need to open fewer facilities while still ensuring that clients are not assigned too far away.

\textbf{Algorithm \textsc{RoundToIntegralFacilities}.} Our next algorithm \textsc{RoundToIntegralFacilities} (Algorithm \ref{alg: round-to-integer-facilities}) rounds the $\Delta$-close fractional solution $(\overline{x}, \overline{y})$ to solution $(x', y')$ with \emph{integral} facilities, but in the process increases the client distances by an additional factor $5$ (i.e., $(x', y')$ is $5\Delta$-close), while not increasing the losses. Our strategy is as follows: we will first find a set of \emph{core clients} $C^* \subseteq C$. Next, we will open a single facility $f(j^*)$ in each $\feas_{j^*}$ for $j^* \in C^*$. This is the integral solution $y^\prime$ for open facilities. These core clients and open facilities satisfy the following two key properties simultaneously:
\begin{enumerate}
    \item For core clients $j^* \in C^*$, the sets $\feas_{j*}$ of feasible facilities have to be mutually disjoint.
    \item For each client $j \in C$, there exists some core client $j^* \in C^*$ such that the open facility $f(j^*)$ is within distance $5\Delta_j$ of $j$.
\end{enumerate}
This is discussed in Section \ref{sec: coreclients}, and these two properties are formalized in Lemma \ref{lem: core-clients}.
We will then assign clients fractionally to $y^\prime$ while maintaining two properties: 
\begin{enumerate}
\item {\it Bounded Losses}: Since sets $\feas_{j^*}$ are disjoint for core clients $j^*$, clients $j$ that are fractionally assigned to any facility in $\feas_{j^*}$ can be reassigned to $f(j^*)$. Thus, the revenue from these clients pays for the facility $f(j^*)$.
\item {\it Bounded Distances}: from the second property above for core clients, each client $j$ can be assigned to some $f(j^*)$ while keeping the objective $g(\cdot)$ within factor $5 \times 4 \max(1, \frac{1}{\delta})$ of the optimal objective.
\end{enumerate}
This is discussed in Section \ref{sec: fractional}. Together, these two steps give us the required integral set of facilities and the corresponding fractional assignments.  

\subsubsection{Finding Core Clients.}\label{sec: coreclients}

\begin{algorithm}[!t]
    \caption{\textsc{CoreClients}$(G, \Delta)$}\label{alg: core-clients}
    \begin{algorithmic}[1]
        \Statex \textbf{input}: an undirected graph $G = (C, E)$ on clients $C$ and a function $\Delta: C \to \R_{\ge 0}$
        \Statex \textbf{output}: $(C^*, \texttt{paths})$ where $C^* \subseteq C$ is a set of \emph{core clients} and $\texttt{paths}_j$ is a path in $G$ from $j$ to some core client $j^* \in C^*$ for each $j \in C$
        \Statex given a path $p = j_0 j_1 \ldots j_T$ for $j_t \in C$ in $G$, define the $\Delta$-length of $p$ as $\Delta(p) = \sum_{t = 0}^T \Delta_{j_t}$
        \State if $G$ is empty, return $(\emptyset, \emptyset)$
        \State choose core client $j^* = {\arg\min}_{j \in C} \Delta_j$, initialize arborescence $\tau$ rooted at $j^*$ and $\texttt{paths}_{j^*} = j^*$
        \State call edge $jj' \in E$ \emph{growing} if $j \not\in \tau$, $j' \in \tau$, and $\Delta_j \ge  \Delta(\texttt{paths}_{j'})$
        \While{there is a growing edge $jj' \in E$}
            \State add directed edge $j' \to j$ to $\tau$
            \State set $\texttt{paths}_{j}$ to be the path from $j$ to $j^*$ in $\tau$
        \EndWhile
        \State obtain $(C^*, \texttt{paths}') = \textsc{CoreClients}(G \setminus \tau, \Delta)$ for the remaining instance
        \State \textbf{return} $(\{j^*\} \cup C^*, \texttt{paths} \cup \texttt{paths}')$
    \end{algorithmic}
\end{algorithm}

For the first step of finding the core set of clients, we present Algorithm \textsc{CoreClients}. For expositional convenience, we first construct the graph $G = (C, E)$ on clients $C$ as follows: there is an edge $jj' \in E$ if and only if $\texttt{feas}_j \cap \texttt{feas}_{j'} \neq \emptyset$, and every vertex (client) $j\in C$ is associated with its $\Delta_j$ value. The $\Delta$-length of a simple path $p = j_0 j_1 \ldots j_T$ in $G$ for clients $j_t \in C$ is defined as the sum $\sum_{t = 0}^T \Delta_{j_t}$ of $\Delta$-values along the path. Algorithm {\sc CoreClients} selects a set of {\it core} clients $C^* \subseteq C$, which form an independent set in the graph $G$, while ensuring a small $\Delta$-length path from any non-core client to some core client. This is formalized in the following lemma: 

\begin{figure}
    \centering
    \begin{minipage}{0.49\linewidth}
        \includegraphics[width=0.9\columnwidth]{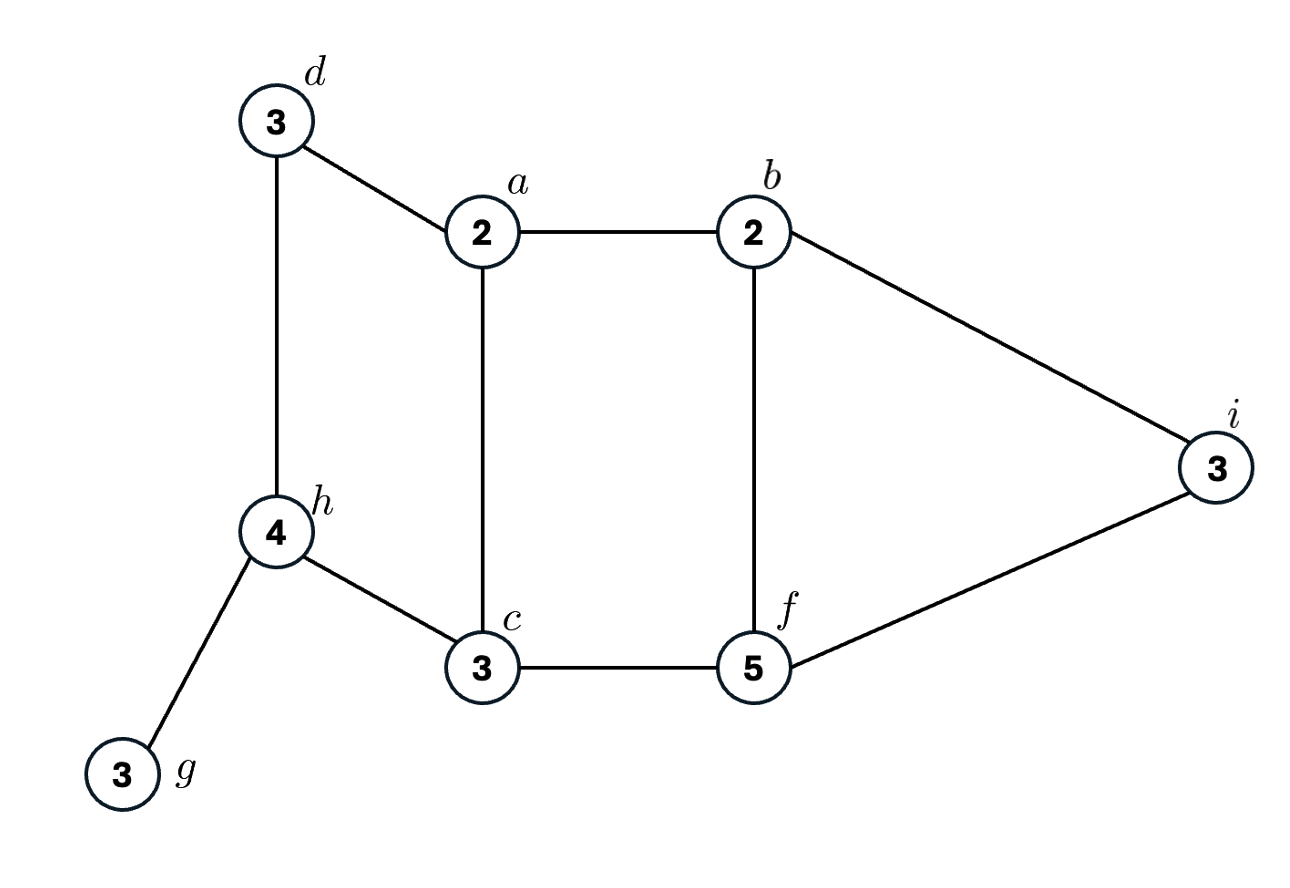}
    \end{minipage}
    \hfill
    \begin{minipage}{0.49\linewidth}
        \includegraphics[width=0.9\columnwidth]{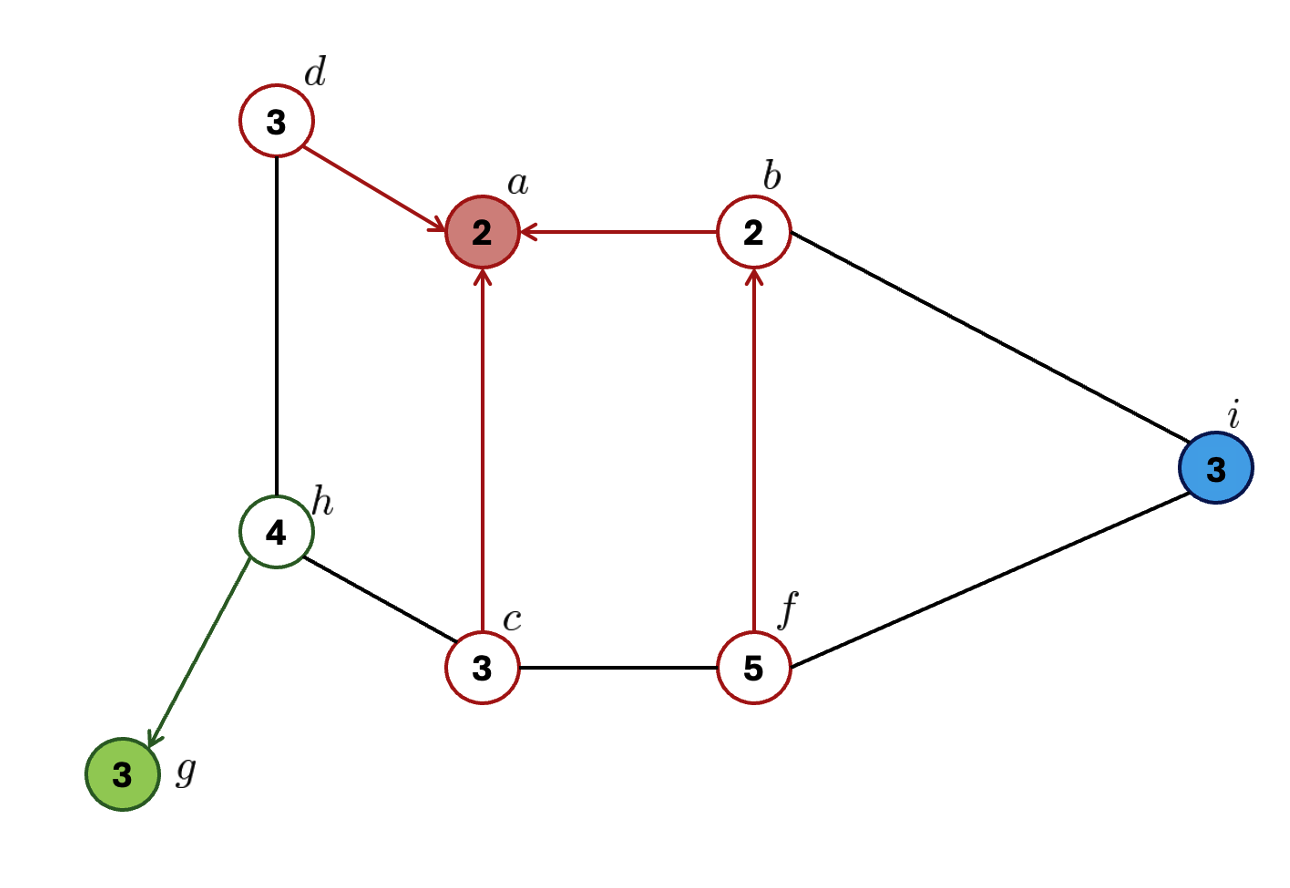}
    \end{minipage}

    \caption{An example to illustrate Algorithm \ref{alg: core-clients}. (left) The graph $G = (C, E)$ with $\Delta$ values for vertices. Initially, the algorithm chooses core client $a = {\arg\min}_{j \in C} \Delta_j$ and forms an arboresence rooted at $a$ on clients $\{a, b, c, d, f\}$. Then, the algorithm chooses core client $g$ and forms the arborescence on clients $\{g, h\}$, and finally, the algorithm chooses the core client $i$. (right) The core clients $C^*$ (shaded) and $\texttt{paths}$, represented through arborescences rooted at the core clients.}
    \label{fig:enter-label}
\end{figure}

\begin{lemma}\label{lem: core-clients}
    Given the graph $G = (C, E)$ on clients $C$ and $\Delta: C \to \R_{\ge 0}$, \textup{\textsc{CoreClients}} finds a subset $C^* \subseteq C$ and  $\textup{\texttt{paths}}_j$ for each $j \in C$ to some $j^* \in C^*$ such that
    \begin{enumerate}
        \item $C^*$ is an independent set of vertices in $G$.
        \item For each client $j \in C$, there is a path $\textup{\texttt{paths}}_j$ from $j$ to some $j^* \in C^*$ with $\Delta$-length at most $\Delta(\textup{\texttt{paths}}_j) \le 2 \Delta_j$.
        
        \item For all neighbors $j \in C$ of a core client $j^* \in C^*$, we have $\Delta_{j} \ge \frac{1}{2} \Delta_{j^*}$.
    \end{enumerate}
\end{lemma}

Before we discuss the proof, at a high level, algorithm \textsc{CoreClients} decomposes the graph $G$ into a set of arborescences that are rooted trees at each core client, while ensuring that each non-core client has a directed path to some core client. This is done by recursively choosing the client $j^*$ with the smallest $\Delta_{j^*}$ value (these are the core clients) and adding all  `nearby' clients to obtain the arborescence rooted at $j^*$. That is, for each client $j$ in this arborescence rooted at $j^*$, the $\Delta$-length of the path from $j$ to $j^*$ is at most $2\Delta_j$.

We then delete the subgraph spanned by the arborescence and recurse on the remaining components of the graph. The complete description is given in Algorithm \ref{alg: core-clients}. We now prove Lemma \ref{lem: core-clients}. 

\emph{Proof.}
1. At any recursive call on a connected component $H$, a core client $j^*$ must have the smallest $\Delta_{j^*}$ value in the current graph $H$. This means that each neighbor $j$ in the current graph must belong to its arborescence, by trivially satisfying $\Delta$-length of the $\Delta(\texttt{paths}_j) = \Delta_j + \Delta_{j^*} \leq 2\Delta_j$. Note that this also implies that any subsequent core client cannot be adjacent to $j^*$.

2. Clearly, the algorithm exhaustively assigns each client $j \in C$ to the arborescence $\tau$ rooted at some core client $j^* \in C^*$. Suppose $j$ was added to $\tau$ through the edge $jj'$. Then, by construction, we have that $\Delta_j \ge \Delta(\texttt{paths}_{j'})$. Since $\texttt{paths}_j$ consists of edge $jj'$ followed by $\texttt{paths}_{j'}$, we have $\Delta(\texttt{paths}_{j'}) = \Delta(\texttt{paths}_{j'}) + \Delta_j \le 2 \Delta_j$.

3. Consider edge $jj^* \in E$ with core client $j^* \in C^*$ and $j \in C$. From Part 1, $j \not\in C^*$. Suppose first that $j$ is in the arborescence of $j^*$. In this case, by Part 2, $\Delta_j + \Delta_{j^*} = \Delta(\texttt{paths}_j) \le 2 \Delta_j$, so that $\Delta_{j^*} \le \Delta_j \le 2 \Delta_j$.

Next, suppose that $j$ is not in the arborescence of $j^*$. Then, $j$ must have been added to the arborescence $\tau$ of some $j_1^*$ chosen in $C^*$ before $j^*$, but this arborescence did not include $j^*$. This can only happen if the $\Delta$-path length $\Delta(\texttt{paths}_{j}) > \Delta_{j^*}$. From Part 2, $2\Delta_j \ge \Delta(\texttt{paths}_{j}) \ge \Delta_{j^*}$, completing the proof. \qed

Note that this lemma gives us our key properties discussed earlier: Part 1 implies that $\feas_{j_1^*} \cap \feas_{j_2^*} = \emptyset$ for all core clients $j_1^*, j_2^* \in C^*$ by definition of the graph $G$. Using Part 2, for all clients $j$, there is some core client $j^*$ such that $\texttt{paths}_{j}$ ends at $j^*$ and therefore $\Delta_{j^*} \le \Delta(\texttt{paths}_j) - \Delta_j \le \Delta_j$, so that $\texttt{dist}_{j, f(j^*)} \le \texttt{dist}_{j, j^*} + \texttt{dist}_{j^*, f(j^*)} \le 4 \Delta_j + \Delta_{j^*} \le 5 \Delta_j$.

\subsubsection{Integral Facilities with Fractional Assignments.}\label{sec: fractional}

\begin{algorithm}[!t]
\caption{\textsc{RoundToIntegralFacilities}$(\overline{x}, \overline{y})$}\label{alg: round-to-integer-facilities}
    \begin{algorithmic}[1]
        \Statex \textbf{input}: $\Delta$-close fractional solution $(\overline{x}, \overline{y})$
        \Statex \textbf{output}: $5\Delta$-close solution $(x', y')$ where $y'$ is integral ($x'$ may be fractional)
        \State for each client $j \in C$, define $\texttt{feas}_j := \{f \in F: \overline{x}_{j, f} > 0\}$
        \State form graph $G = (C, E)$ on vertex set $C$ with edge $j_1 j_2 \in E$ if and only $\texttt{feas}_{j_1} \cap \texttt{feas}_{j_2} \neq \emptyset$
        \State $(C^*, \texttt{paths}) \gets \textsc{CoreClients}(G, \Delta)$
        \Statex \textbf{Phase 1: open facilities and partially assign clients}
        \For{$j^* \in C^*$}
            \State let $f(j^*) := {\arg\min}_{f \in \texttt{feas}_{j^*}} c_f$ be the cheapest facility in $\texttt{feas}_{j^*}$ \label{step: open-cheapest-facility}
            \State set $y'_{f(j*)} = 1$,  and assign $x_{j^*, f(j^*)} = 1$ \Comment{Open facility and assign}
            \For{each neighbor $j$ of $j^*$ in $G$}
                \State \label{step: client-assignments-integral-facilities} set $x'_{j, f(j^*)} = \sum_{f \in \texttt{feas}_{j^*}} \overline{x}_{j, f}$ to be the contribution of $j$ to facilities in $\texttt{feas}_{j^*}$
            \EndFor
        \EndFor
        \Statex \textbf{Phase 2: fully assign clients}
        \For{each client $j \in C$ that is partially assigned, i.e., $\sum_{f} x'_{j, f} < 1$}
            \State let $j^* \in C^*$ be the endpoint of path $\texttt{paths}(j)$
            \State increase $x'_{j, f(j^*)} \gets x'_{j, f(j^*)} + \left(1 - \sum_{f} x'_{j, f}\right)$ \color{black} \Comment{Fully assign $j$}
        \EndFor
        \State set any undefined $x'_{j, f}$ and $y'_f$ to $0$
        \State \textbf{return} $(x', y')$
    \end{algorithmic}
\end{algorithm}

Taking the $\Delta$-close fractional solution $(\overline{x}, \overline{y})$, we will return a $5\Delta$-close solution $(x^\prime, y^\prime)$ with $y^\prime$ integral using the following Algorithm  \textsc{RoundToIntegralFacilities} (Alg. \ref{alg: round-to-integer-facilities}). Using \textsc{CoreClients} as a subroutine, 
the algorithm runs in two phases, where Phase 1 opens the cheapest facility $f(j^*) \in \feas_{j^*}$ for each core client $ j^* \in C^*$ and partially assigns clients to bound the losses of these facilities as described above. Each client's contribution to any facility in $\feas_{j^*}$ is assigned to the chosen $f(j^*)$. Phase 2 then assigns the remainder contribution of the clients to the open facility $f(j^*)$ where $j^*$ is the end of the path $\texttt{paths}_j$ obtained using $\textsc{CoreClients}$. This keeps their distances to assigned facilities bounded. 

In Phase 1, for each core client $j^* \in C^*$, the cheapest facility $f(j^*) \in C^*$ is opened. For each neighbor $j$ of $j^*$ in $G$, we fractionally assign $j$ to $f(j^*)$, setting $x'_{j, f(j^*)}$ to be the contribution of $j$ to facilities in $\feas_{j^*}$, that is, $x'_{j, f(j^*)} = \sum_{f \in \feas_{j^*}} \overline{x}_{j, f}$. First, we show that \textsc{RoundToIntegralFacilities} returns a valid fractional solution $(x', y')$. It is clear that $(x', y')$  satisfies constraints (\ref{cons: assign-clients-only-to-open-facilities}) in the relaxation of (\ref{eqn: price-of-fairness-ip}).
It is sufficient to show that $\sum_{f} x'_{j, f} \le 1$ for all $j \in C$, since in Phase 2 (line 11) we ensure that $\sum_f x'_{j, f} \ge 1$ for all $j \in C$.

\begin{claim}
    For each client $j \in C$, $\sum_{f} x'_{j, f} \le 1$ at the end of Phase 1 in Alg. \textup{\textsc{RoundToIntegralFacilities}}.
\end{claim}

\emph{Proof.} $\texttt{feas}_{j_1^*}$ and $\texttt{feas}_{j_2^*}$ are disjoint for all distinct core clients $j_1^*, j_2^* \in C^*$. Therefore, at the end of Phase 1, $\sum_f x'_{j, f} = \sum_{j^* \in C^*} x'_{j, f(j^*)} = \sum_{j^* \in C^*} \sum_{f \in \texttt{feas}_{j^*}} \overline{x}_{j, f} \le \sum_{f \in \texttt{feas}_j} \overline{x}_{j, f} = 1.$ \qed

Our next lemma bounds the increase in client distances:

\begin{lemma}\label{lem: integral-facilities-bounded-distances}
    For each client $j \in C$ and facility $f \in F$, if $x'_{j, f} > 0$ then $\textup{\texttt{dist}}_{j, f} \le 5 \Delta_j$.
\end{lemma}

\emph{Proof.}
    Each client $j \in C$ is potentially (fractionally) assigned to some facilities in Phase 1 and to at most one facility in Phase 2.

    \textbf{Phase 1.} Suppose $j$ is fractionally assigned to some $f(j^*)$ for $j^* \in C^*$ in Phase 1. Then $j$ is a neighbor of $j^*$, so there is some $f \in \feas_{j} \cap \feas_{j^*}$, that is, $\overline{x}_{j, f} > 0$ and $\overline{x}_{j^*, f} > 0$. Algorithm \textsc{CoreClients} guarantees that $\Delta_j \ge \frac{1}{2} \Delta_{j^*}$ in this case (Lemma \ref{lem: core-clients}), so that
    \begin{align*}
        \texttt{dist}_{j, f(j^*)} &\le \texttt{dist}_{j, f} + \texttt{dist}_{j^*, f} + \texttt{dist}_{j^*, f(j^*)} \\
        &\le \Delta_j + \Delta_{j^*} + \Delta_{j^*} \le \Delta_j + 2 \Delta_j + 2 \Delta_j = 5 \Delta_j.
    \end{align*}
    
    \textbf{Phase 2.} $j$ is assigned to $f(j_T)$ for $j_T \in C^*$ where $\texttt{paths}(j) = j j_1 j_2 \ldots j_{T - 1} j_T$ is a path in $G$. Algorithm \textsc{CoreClients} guarantees (Lemma \ref{lem: core-clients}) that $\Delta_{j_1} + \Delta_{j_2} + \ldots + \Delta_{j_T} \le \Delta_j$, and therefore the distance of $j$ to $f(j_T)$ is at most
    \[
        2(\Delta_{j_1} + \Delta_{j_2} + \ldots + \Delta_{j_T}) + \Delta_{j} \le 3 \Delta_{j}. \quad \textup{\qed}
    \]

Our next lemma shows that the total loss of unprofitable facilities in $(x', y')$ is at most the total loss of unprofitable facilities in $(\overline{x}, \overline{y})$:

\begin{lemma}\label{lem: integral-facility-bounded-loss}
    The total loss for the rounded solution $(x', y')$ returned by \textup{\textsc{RoundToIntegralFacilities}}$(\overline{x}, \overline{y})$ is $\sum_{f \in F} \loss'_f \le \sum_{f \in F} \overline{\loss_f}$.
\end{lemma}

The proof is based on the following idea: we only open one facility $f(j^*)$ for each core client $j^* \in C^*$. Further, the set $C^*$ is an independent set in the graph $G = (C, E)$. Recall that for each $j^* \in C^*$, its neighbors in $G$ contribute some revenue to $f(j^*)$ in Phase 1 of the algorithm. We show that since $f(j^*)$ is chosen as the cheapest facility in $\feas_{j^*}$, this contribution from the neighbors of $j^*$ is sufficient to offset the operating cost of $f(j^*)$, up to the loss incurred in the fractional solution $(\overline{x}, \overline{y})$. We defer the proof of this lemma to Appendix \ref{app: integral-facilities-bounded-loss-proof}.

Note that since the total loss $\sum_{f} \overline{\loss}_f$ of $(\overline{x}, \overline{y})$ is at most $2\delta \sum_{j \in C} r_j$ (Lemma \ref{lem: alpha-point-rounding}), we have from the above lemma that $\sum_{f} \loss'_f \le 2\delta \sum_{j \in C} r_j$ as well.  Now that we have a solution $(x', y')$ with integral open facilities $\{f \in F: y'_f = 1\}$, it remains to round the fractional assignments $x'$.

\subsection{Finding Integral Assignments}\label{sec: rounding-last-step}

In our final rounding step, we find integral assignments for the $2\delta$-subsidized fractional solution $(x', y')$ with integral $y'$ and subsidy at most $2\delta$. This rounding procedure gives an integral solution $(x'', y')$ but increases the subsidy to at most $2\delta + \theta$. This proof crucially uses the $\theta$-small revenues assumption (see Section \ref{sec: intro-fsfl}), i.e., $r_j \le \theta c_f$ for all clients $j \in C$ and facilities $f \in F$. We have the following more general lemma:

\begin{lemma}\label{lem: integral-assignment}
    There exists a polynomial-time rounding algorithm that given a $\delta'$-subsidized $\Delta'$-close fractional solution $(x', y')$ where $y'$ is integral, obtains a $(\delta' + \theta)$-subsidized $\Delta'$-close integral solution $(x'', y')$ for any instance of FSFL that satisfies the $\theta$-small revenues assumption for given $\theta > 0$.
\end{lemma}

To prove this lemma, we consider a parallel scheduling problem where the machines are open facilities $F' = \{f \in F: y'_f = 1\}$, jobs are clients $C$, and processing time $p_{j, f} = r_j$ if $x'_{j, f} > 0$ and $p_{j, f} = \infty$ otherwise (i.e., job $j$ cannot be assigned to machine $f$). Denote the set of clients that can be assigned to $f$ as $C_f := \{j \in C: x'_{j, f} > 0\}$.  Our proof uses \cite{shmoys_approximation_1993}'s scheduling algorithm as a subroutine, stated in terms of our problem:
\begin{lemma}[\cite{shmoys_approximation_1993}]\label{lem: generalized-assignment}
    There exists a polynomial-time algorithm that given a fractional schedule $x'$, returns an integral schedule $x''$ where (1) each job/client $j$ is assigned to exactly one facility/machine in the support of $x'$ and (2) for each machine $f \in F'$, the total load $\sum_{j \in C_f} x''_{j, f} r_j$ on $f$ under $x''$ is at most the total load $\sum_{j \in C_f} x'_{j, f} r_j$ on $f$ under $x'$, plus the load $r_{j(f)}$ of at most one extra job $j(f) \in C_f$, i.e.,
    \[
        \sum_{j \in C_f} x''_{j, f} r_j \le r_{j(f)} + \sum_{j \in C_f} x'_{j, f} r_j.
    \]
\end{lemma}
    
Lemma \ref{lem: generalized-assignment} allows us to round the fractional assignment to an integer assignment, while adding revenue from a single client to each facility. However, to bound the total loss in this assignment to prove Lemma \ref{lem: integral-assignment}, one needs additional algebraic arguments, which we include in Appendix \ref{app: integral-assignment}.

With Lemma \ref{lem: integral-assignment} in hand, we are ready to complete the proof of our main Theorem \ref{thm: price-of-fairness} that gives the bicriteria oracle for FSFL:

\emph{Proof of Theorem \ref{thm: price-of-fairness}}. Denote the optimal (integral) $\delta$-subsidized solution as $(x^*, y^*)$ with objective value $g(d^*)$. Since $(x, y)$ is the optimal fractional solution, (1) it must be $\delta$-subsidized and (2) the objective value $g(d) \le g(d^*)$, where $d \in \R^C$ is the vector of client distances for $(x, y)$.

The solution $(\overline{x}, \overline{y})$ returned by $\alpha$-\textsc{PointRounding} is $\left(4 \max \left(1, \frac{1}{\delta}\right) d\right)$-close by Lemma \ref{lem: alpha-point-rounding}.1. The solutions $(x', y')$ and $(x'', y')$ that give integral facilities and integral assignment respectively are both $\left(5 \times 4 \cdot \max \left(1, \frac{1}{\delta}\right) d\right)$-close by Lemmas \ref{lem: integral-facilities-bounded-distances} and \ref{lem: integral-assignment} respectively. Therefore, since $g$ is sublinear, for distance vector $d''$ for integral solution $(x'', y')$,
\[
    g(d'') = O\left(\max \left(1, \frac{1}{\delta}\right)\right) g(d) = O\left(\max \left(1, \frac{1}{\delta}\right)\right) g(d^*).
\]

The solution $(\overline{x}, \overline{y})$ returned by $\alpha$-\textsc{PointRounding} is $2\delta$-subsidized by Lemma \ref{lem: alpha-point-rounding}.2. The solutions $(x', y')$ returned by \textsc{RoundToIntegralFacilities} is also $2\delta$-subsidized by Lemma \ref{lem: integral-facility-bounded-loss}, and the solution $(x'', y')$ is $(2\delta + \theta)$-subsidized by Lemma \ref{lem: integral-assignment}. \qed

To summarize, we presented a bicriteria oracle for approximating the facility location with subsidies, by step-wise rounding the fractional optimal solution while ensuring that the total losses are bounded by a fraction of the revenue and that the client distances do not blow up by too much.

\subsection{Portfolios for Fair Subsidized Facility Location}\label{sec: ffl-portfolios}

In this section, we obtain polynomial-time portfolios for FSFL for objectives in  $\class = L_p$ norms of group distances. Using Theorem \ref{thm: portfolios-lp-norms}, we can reduce the problem of obtaining portfolios to the problem of designing approximation oracles, and using Theorem \ref{thm: price-of-fairness}, we get a bicriteria oracle for FSFL, as discussed in the previous section. Putting these together, we get:

\portfolioFFL*

Note that the portfolio size grows logarithmically in the number of client groups $t$. Therefore, even with intersectional groups, the number of solutions in the portfolio grows slowly. For example, if we have 2 categories for urbanization, 2 categories for income levels, and 4 categories for geographical region, this still results in only $t = 16$ groups. Similarly, for $\delta < 1$, the dependence on $1/\delta$ is logarithmic, and for $\delta \ge 1$ the portfolio size is independent of $\delta$. Further, these are theoretical worst-case bounds, and in practice, the portfolio sizes are much smaller (see Section \ref{sec: experiments}).

Next, we present a lower bound for portfolios for FSFL, which shows that our results are order-optimal. That is, one cannot hope to construct a smaller O(1)-approximate portfolio for this problem. 

\portfolioFFLLowerBound*

The main idea of the proof is as follows: for all large enough $t$ and for all constants $\alpha > 1$, we give an instance of FSFL where any $\alpha$-approximate portfolio size must be $\simeq  \log_{2\alpha} t$. Specifically, we construct an instance with exactly $\simeq  \log_{2\alpha} t$ distinct feasible solutions. Further, we give a set of $\simeq  \log_{2\alpha} t$ norms such that for each fixed $L_p$ norm in this set, exactly one of the $\simeq  \log_{2\alpha} t$ solutions is optimal and every other solution is not an $\alpha$-approximation. Thus, any portfolio for all $L_p$ norms must contain each of the $\simeq \log_{2\alpha} t$ solutions. We include the details in Appendix \ref{app: fsfl-portfolio-lower-bound}.

\section{Experiments}\label{sec: experiments}

We now present our experiments on U.S. Census and pharmacy data in the state of Mississippi. For different values of the subsidy parameter $\delta$, our goal is to present a portfolio of solutions to open $10$ new pharmacies in Mississippi to help tackle the issue of medical deserts in an equitable way. Recall that (see Figure \ref{fig: online-tool}) we define a medical desert as a Census blockgroup\footnote{A Census blockgroup is a small administrative region with between 500 and 3000 people. It is the smallest unit for which Census data is publicly available.} with over $20\%$ below the poverty line and at a distance of over $2$ miles (urban areas) or $10$ miles (rural areas) from its nearest CVS/Walgreens/Walmart pharmacy.\footnote{We choose a higher distance threshold for rural areas to account for better access to vehicles. This is in line with the methodology of the U.S. government to document food deserts \citep{usda_usda_2023}.} Focusing on these three largest pharmacy chains allows us to understand the impact of opening multiple facilities at scale.

As Figure \ref{fig: online-tool} notes, 348 of the 2445 blockgroups in Mississippi are medical deserts, and they disproportionately affect the majority Black or African American population. Using our integer programming formulation (\ref{eqn: price-of-fairness-ip}) for FSFL and our portfolio algorithm for $L_p$ norms from Section \ref{sec: monotonically-interpolating-portfolio}, we will give a portfolio of solutions for various values of the subsidy parameter $\delta$. Each solution in the portfolio recommends the locations of $10$ new pharmacies in Mississippi in addition to the 206 existing facilities. The total loss of these new pharmacies is bounded by a fraction $\delta$ of the total revenue of all clients.

We choose $\delta \in \{0.005, 0.01, 0.02, 0.05\}$ to allow the losses to span from $0.5\%$ to $5\%$. Each blockgroup is a client, and a facility can be opened at any blockgroup location, where the location of a blockgroup is taken to be its geometric center. Further details on modeling choices can be found in Appendix \ref{app: further-details-on-experiments}.

We classify each blockgroup into $4 \times 2 \times 2 = 16$ groups based on (1) which of the 4 Congressional districts in Mississippi it lies in, (2) whether the blockgroup is rural or urban, and (3) whether or not $20\%$ of the people in the blockgroup are below the poverty line. The group distance for a group is defined as the average distance of blockgroups to facilities, weighted by the population of the blockgroup. Distances of urban groups are weighted five times as much as rural groups to account for lesser access to vehicles. Given $p \ge 1$, the $L_p$ norm objective minimizes the $L_p$ norm of the $16$-dimensional group distance vector.

We set the approximation parameter $\varepsilon$ in Theorem \ref{thm: portfolios-lp-norms} to $0.15$, so that we are guaranteed a $1.15$-approximate portfolio for all $L_p$ norm objectives since we obtain exact solutions to the \ref{eqn: price-of-fairness-ip}.

\begin{figure}
    \begin{minipage}{0.75\columnwidth}
                    \centering
        \includegraphics[width=\linewidth]{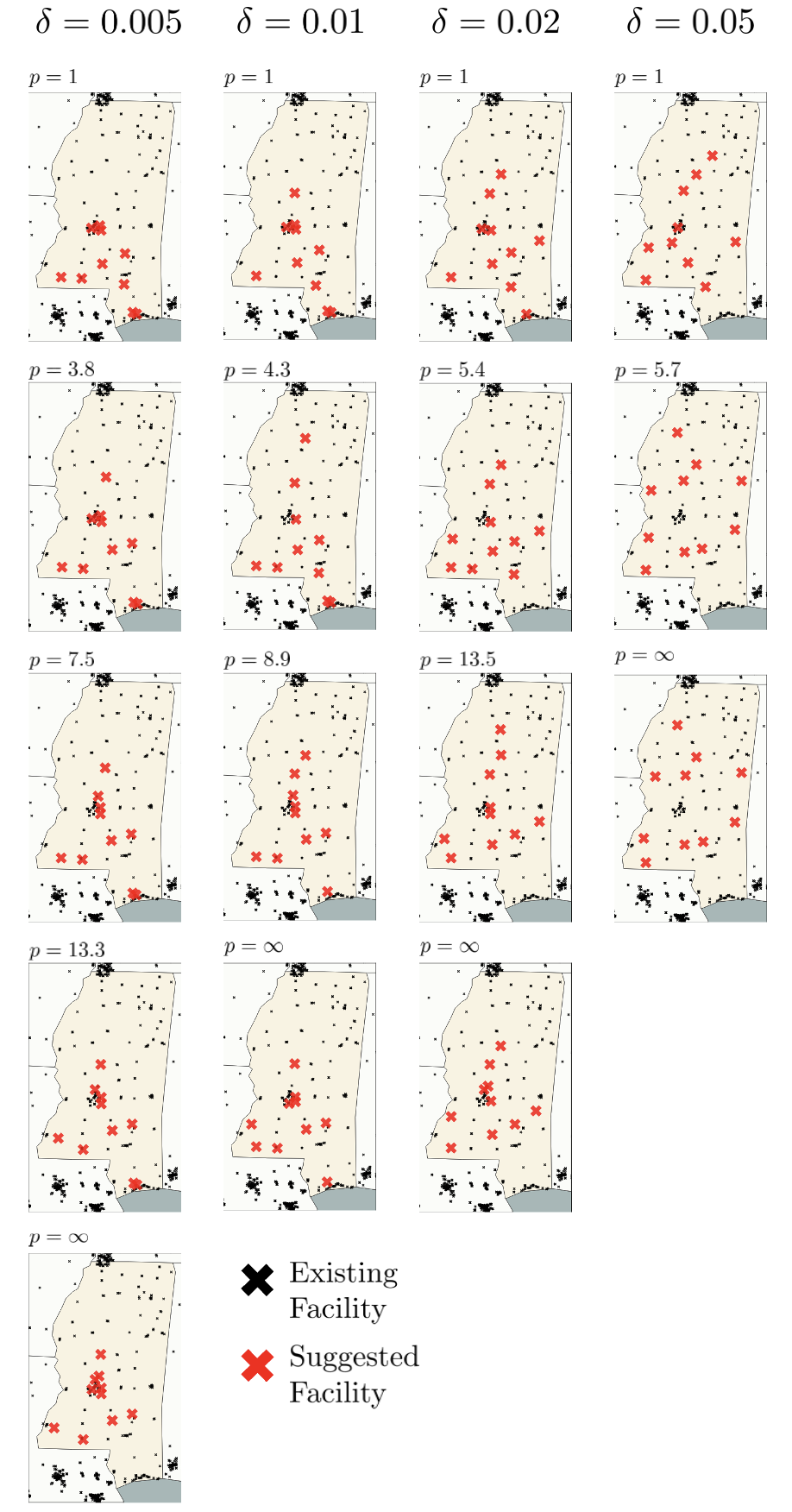}
    \end{minipage}
    \hfill
    \begin{minipage}{0.24\textwidth}

        \caption{Portfolios of suggested locations for $k = 10$ new pharmacies using the FSFL model in the state of Mississippi, USA, in addition to existing CVS, Walmart, and Walgreens pharmacies. Each column shows the portfolio for a given subsidy parameter $\delta \in \{0.005, 0.01, 0.02, 0.05\}$ for approximation factor $\alpha = 1.15$. While different solutions recommend opening facilities in different locations, all solutions significantly reduce the number of medical deserts.}
        \label{fig: new-facility-locations}
    \end{minipage}
\end{figure}

\textbf{Results.} Figure \ref{fig: new-facility-locations} shows the portfolio for each value of $\delta \in \{0.005, 0.01, 0.02, 0.05\}$. For example, the portfolio for $\delta = 0.02$ has four solutions corresponding to $p = 1, 5.4, 13.5, \infty$ arranged in the third column. We further expand on our results for this portfolio for $\delta = 0.02$.

As Table \ref{tab: number-of-medical-deserts} shows, \emph{each solution in this portfolio reduces between 43 and 51 medical deserts out of 348 medical deserts}, despite adding only 10 facilities each to the existing 206 facilities. Further, \emph{70-80\% of these $\sim$50 blockgroups are African American for each solution, thus significantly reducing the disproportionate impact of medical deserts on the Black population.}

The portfolio solutions are also significantly diverse across the 16 groups. As Tables \ref{tab: spider-plots} and \ref{tab: number-of-medical-deserts} and Figure \ref{fig: new-facility-locations} show, \emph{different solutions are optimal for different groups in terms of reduction of distances or medical deserts, with each solution performing well for different groups}. Each solution has its own strength: the $L_1$ norm solution reduces the most number of medical deserts, the $L_{5.4}$ norm solution leads to the smallest average distance traveled, and the $L_\infty$ norm solution is the most equitable. Therefore, this work helps move away from specific model formulations and highlights the trade-offs of a small set of solutions for various communities. We will make our tool publicly available, and the assumptions on cost and revenue can be modified by decision-makers to suggest a small portfolio of options.

\begin{table}[]
\begin{minipage}{0.65\textwidth}
    \centering\small
\begin{tabular}{|cc||c||cccc|}
\hline
\multicolumn{2}{|c||}{Group} &
  \multirow{2}{*}{\begin{tabular}[c]{@{}c@{}}Existing\\ Medical\\ Deserts\end{tabular}} &
  \multicolumn{4}{c|}{Number of medical deserts reduced} \\ \cline{1-2} \cline{4-7} 
\multicolumn{1}{|c|}{\begin{tabular}[c]{@{}c@{}}Urban/\\ Rural\end{tabular}} &
  District: &
   &
  \multicolumn{1}{c|}{\begin{tabular}[c]{@{}c@{}}$L_1$\\ Norm\end{tabular}} &
  \multicolumn{1}{c|}{\begin{tabular}[c]{@{}c@{}}$L_{5.4}$\\ Norm\end{tabular}} &
  \multicolumn{1}{c|}{\begin{tabular}[c]{@{}c@{}}$L_{13.5}$\\ Norm\end{tabular}} &
  \begin{tabular}[c]{@{}c@{}}$L_{\infty}$\\ Norm\end{tabular} \\ \hline
\multicolumn{1}{|c|}{\multirow{4}{*}{Rural}} &
  1 &
  41 &
  \multicolumn{1}{c|}{0} &
  \multicolumn{1}{c|}{0} &
  \multicolumn{1}{c|}{0} &
  0 \\ \cline{2-7} 
\multicolumn{1}{|c|}{} &
  2 &
  109 &
  \multicolumn{1}{c|}{16} &
  \multicolumn{1}{c|}{22} &
  \multicolumn{1}{c|}{\textbf{23}} &
  \textbf{23} \\ \cline{2-7} 
\multicolumn{1}{|c|}{} &
  3 &
  98 &
  \multicolumn{1}{c|}{19} &
  \multicolumn{1}{c|}{19} &
  \multicolumn{1}{c|}{\textbf{20}} &
  19 \\ \cline{2-7} 
\multicolumn{1}{|c|}{} &
  4 &
  26 &
  \multicolumn{1}{c|}{\textbf{6}} &
  \multicolumn{1}{c|}{5} &
  \multicolumn{1}{c|}{0} &
  0 \\ \hline
\multicolumn{1}{|c|}{\multirow{4}{*}{Urban}} &
  1 &
  6 &
  \multicolumn{1}{c|}{0} &
  \multicolumn{1}{c|}{0} &
  \multicolumn{1}{c|}{0} &
  0 \\ \cline{2-7} 
\multicolumn{1}{|c|}{} &
  2 &
  31 &
  \multicolumn{1}{c|}{\textbf{4}} &
  \multicolumn{1}{c|}{0} &
  \multicolumn{1}{c|}{0} &
  \textbf{4} \\ \cline{2-7} 
\multicolumn{1}{|c|}{} &
  3 &
  26 &
  \multicolumn{1}{c|}{\textbf{1}} &
  \multicolumn{1}{c|}{\textbf{1}} &
  \multicolumn{1}{c|}{0} &
  0 \\ \cline{2-7} 
\multicolumn{1}{|c|}{} &
  4 &
  11 &
  \multicolumn{1}{c|}{\textbf{5}} &
  \multicolumn{1}{c|}{0} &
  \multicolumn{1}{c|}{0} &
  0 \\ \hline \hline
\multicolumn{2}{|c||}{Total} &
  348 &
  \multicolumn{1}{c|}{51} &
  \multicolumn{1}{c|}{47} &
  \multicolumn{1}{c|}{43} &
  46 \\ \hline
\end{tabular}
\end{minipage}
\hfill
\begin{minipage}{0.31\textwidth}
    \caption{Reduction in the number of medical deserts for different solutions in the portfolio for subsidy $\delta = 0.02$ for various groups in Mississippi. A medical desert is a blockgroup with $\ge 20\%$ poverty rate and over $n$ miles further away from the nearest pharmacy chain, where $n = 2$ miles for urban areas and $n = 10$ miles for rural areas.}
    \label{tab: number-of-medical-deserts}
\end{minipage}
\end{table}

\section{Conclusion}\label{sec: conclusion}

We introduced the notion of portfolios of solutions for optimization problems with a (potentially infinite) class of objectives, which are a small number of feasible solutions to the problem that approximately optimize each objective in the class. We gave trade-offs between portfolio size and approximation quality for the classes of (1) conic combinations of given base functions and (2) functions that interpolate monotonically between the sum and max of the base functions. In particular, for the latter, we guarantee a portfolio size at most logarithmic in the number of base functions. Our algorithms reduce the problem of finding portfolios to the problem of designing approximation algorithm oracles for fixed objectives.

As a concrete application, we proposed the Fair Subsidized Facility Location model motivated by the crisis of pharmacy deserts. Our model is a generalization of the classical facility location problems and allows for richer clients to contribute to new facilities in underserved areas through a `subsidy' constraint. It also allows the losses to be controlled while simultaneously being fair to different groups of people. This is the first paper to give a novel approximation algorithm for FSFL, introducing a new combinatorial subroutine to round fractional solutions to get an integral set of facilities. However, it remains open if the dependence of the approximation ratio on $1/\delta$ is necessary.

We used our FSFL model to propose portfolios for opening new pharmacies in Mississippi, USA, with the aim of reducing medical deserts and average distances to the nearest pharmacy. Solutions in our portfolio not only significantly reduce the number of medical deserts using only 10 new facilities each, they are also diverse in their effects on people living in different areas.
Thus, a policymaker can evaluate these solutions on multiple axes of interest and weigh their relative trade-offs before making an informed decision, as opposed to relying on the algorithm designer for a single good solution. Further, these options may also help constituents in different districts to mobilize political support for facilities opening closer to them.

Portfolios can serve as a general tool for policymaking that is useful beyond the specific application of pharmacy deserts, and provide a practical but mathematically robust way for decision-makers to make informed decisions. One limitation of our approach (by design) is the requirement of specifying objectives that the decision-makers care about, e.g., $\class = L_p$ norms of group distances. It would be interesting to consider formulations where the set of objectives is not completely specified or is uncertain. Further, in the FSFL problem, this work assumes that client groups are given \emph{a priori}, and it is unclear how sensitive the portfolio solutions are to the choice of client groups (and if their group membership might be private, or if they can misreport). We leave such explorations of stochastic or adaptive portfolios to future work. 

\bibliographystyle{plain}
\bibliography{references,references-zotero}

\newpage

\appendix

\section{Portfolio Size Lower Bound for Conic Combinations}\label{app: convex-combination-portfolio-lower-bound}

We show that the portfolio size for conic combinations $\class = \{\sum_{j \in [N]} \lambda_j h_j: \lambda \ge 0\}$ of base functions $h_1, \ldots, h_N$ must be exponential in $N$ in some cases. In fact, our base functions $h_i, i \in [N]$ will be linear.

\begin{lemma}
    For all $N \ge 1$, there exist $\mathcal{D} \subseteq \R_{\ge 0}^{N}$ and base functions $h_1, \ldots, h_N$ such that any $2$-approximate portfolio for the class $\class = \{\sum_{j \in [N]} \lambda_j h_j: \lambda \ge 0\}$ of conic combinations of base functions must have size $\ge 2^{N} - 1$.
\end{lemma}

\emph{Proof.}
    We specify the base functions first: $h_i(x) = x_i$ for all $x \in \R^{N}$ and $i \in [N]$ (i.e., the $i$th coordinate). Given a set $S \subseteq [N]$, let $\chi_S \in \{0, 1\}^N$ denote its characteristic vector. Then for all $S \neq \emptyset$, the function $h_S(x) := \chi_S^\top x \in \class$. We will construct $\mathcal{D} = \{x(S): S \neq \emptyset\}$ with $|\mathcal{D}| = 2^{N} - 1$ such that for each $S$, there will be a unique minimizer $x(S) = \min_{x \in \mathcal{D}}h_S(x)$ and further that for all $T \neq S$, $x(T)$ will not be a $2$-approximation for $h_S$.

    Fix constants $a = 3$ and $b = 2N a^{N}$ that we will use to define the vectors $x(S)$. For all $T \neq \emptyset$, define $x(T)$ as follows:
    \[
        x(T)_i = \begin{cases}
            a^{|T|} & \text{if} \ i \in T, \\
            b & \text{if} \ i \not\in T.
        \end{cases}
    \]

    Given some $S, T \neq \emptyset$, we get that
    \[
        h_S(x(T)) = \sum_{i \in S \cap T} a^{|T|} + \sum_{i \in S \setminus T} b = |S \cap T| a^{|T|} + b |S \setminus T|. 
    \]
    In particular, $h_S(x(S)) = |S| a^{|S|}$. We will show that for all $T \neq S$, $h_S(x(T)) > 2 h_S(x(S))$.

    \textbf{Case I: $S \subsetneq T$}. Then $h_S(x(T)) = |S| a^{|T|}$. Since $S \neq T$, we must have $|T| > |S|$ and so $h_S(x(T)) > a \cdot |S| a^{|S|} = 3 \cdot h_S(x(S))$.

    \textbf{Case II: $S \setminus T \neq \emptyset$}. Then $h_S(x(T)) > b = 2N a^{N} \ge 2|S| a^{|S|} = 2 h_S(x(S))$. This completes the proof. \qed

\section{Omitted Algorithms and Proofs for FSFL}\label{sec: omitted-proofs-algorithm}

We present omitted algorithms and proofs for the rounding algorithm for FSFL in Section \ref{sec: approximation-algorithm} here. First, we present algorithm $\alpha$-\textsc{PointRounding} and its proof in Appendix \ref{sec: alpha-point-rounding}.
Lemma \ref{lem: integral-facility-bounded-loss}, which bounds the subsidy of \textsc{RoundToIntegralFacilities}, is proven in Appendix \ref{app: integral-facilities-bounded-loss-proof}. Proof of Lemma \ref{lem: integral-assignment} that gives integral assignments given integral facilities is given in Appendix \ref{app: integral-assignment}.

\subsection{Algorithm $\alpha$-\textsc{PointRounding}}\label{sec: alpha-point-rounding}

We give the $\alpha$-point rounding algorithm from \cite{STA97} for completeness in Algorithm \ref{alg: alpha-point-rounding}. We restate the guarantee of the algorithm here:

\begin{algorithm}
    \caption{\textsc{$\alpha$-PointRounding}$(x, y)$}\label{alg: alpha-point-rounding}
    \begin{algorithmic}[1]
        \Statex \textbf{input}: fractional solution $(x, y)$
        \Statex \textbf{output}: fractional solution $(\overline{x}, \overline{y})$
        \State set $\alpha = \frac{\delta + 2}{2(\delta + 1)}$
        \For{each client $j \in C$}
            \State find order $\pi$ on facilities in $F$ so that $\texttt{dist}_{j, \pi(1)} \le \texttt{dist}_{j, \pi(2)} \le \ldots $ 
            \State for indices $k = 1, 2, \ldots$ denote the prefix sum $z_{j, k} := \sum_{i \le k} x_{j, \pi(i)}$
            \State let $k_j$ be the smallest index such that $z_{j, k_j} \ge \alpha$
            \State set
            \[
                \overline{x}_{j, \pi(i)} = \begin{cases}
                    \frac{1}{z_{j, k_j}} \cdot x_{j, \pi(i)} & \text{if} \ i \le k_j, \\
                    0 & \text{if} \ i > k_j.
                \end{cases}
            \]
        \EndFor
        \State for each facility $f \in F$, set $\overline{y}_f = \frac{1}{\alpha} y_f$ \\
        \Return $(\overline{x}, \overline{y})$
    \end{algorithmic}
\end{algorithm}

\alphaPointRounding*

To prove this lemma, we first need another result for $\alpha$-\textsc{PointRounding}, which states that the distribution of client assignment to various facilities does not change significantly in $\alpha$-\textsc{PointRounding}:

\begin{lemma}\label{lem: alpha-point-rounding-mass-shift}
    For all clients $j \in C$ and all facility subsets $F_1 \subseteq F$,
    \[
        \sum_{f \in F_1} \overline{x}_{j, f} \ge - (1 - \alpha) + \sum_{f \in F_1} x_{j, f}.
    \]
\end{lemma}

\emph{Proof.}
    Note that we have $\sum_{f \in F} x_{j, f} = \sum_{f \in F} \overline{x}_{j, f} = 1$ along with nonnegativity of $x, \overline{x}$, and therefore $\alpha$-\textsc{PointRounding} can be viewed as shifting some mass from the prior distribution $x_j$ to the posterior distribution $\overline{x}_j$. Intuitively, the lemma says that the algorithm can shift no more than fraction $1 - \alpha$ of the mass away from any subset $F_1$ of facilities.
    
    To see this, we use some notation from the algorithm: $\pi$ is the ordering on facilities such that $\texttt{dist}_{j, \pi(1)} \le \texttt{dist}_{j, \pi(2)} \le \ldots$, and $k_j$ is the least index $k$ such that $\sum_{i \le k} x_{j, \pi(i)} \ge \alpha$ Call facility $\pi(i)$ \emph{massive} if $i \le k_j$ and \emph{light} otherwise. Therefore the total prior mass of light facilities is at most $1 - \sum_{i \le k} x_{j, \pi(i)} \le 1 - \alpha$. By construction, $\overline{x}_{j, \pi(i)} < x_{j, \pi(i)}$ if and only if $i$ is light. Therefore, for any $F_1 \subseteq F$, at most $1 - \alpha$ mass can be reduced from it, i.e.,
    \[
        \sum_{f \in F_1} x_{j, f} - \sum_{f \in F_1} \overline{x}_{j, f} \le 1 - \alpha. 
    \]
    {\flushright \qed}

\emph{Proof of Lemma \ref{lem: alpha-point-rounding}.}
    \begin{enumerate}
        \item This part follows the argument from  \cite{STA97}. We present the proof here for completeness. For any client $j \in C$, by definition,
        \begin{align*}
            \dist_j &= \sum_{i} \texttt{dist}_{j, \pi(i)} x_{j, \pi(i)} = \sum_{i < k_j} \texttt{dist}_{j, \pi(i)} x_{j, \pi(i)} + \sum_{i \ge k_j} \texttt{dist}_{j, \pi(i)} x_{j, \pi(i)} \\
            &\ge \sum_{i \ge k_j} \texttt{dist}_{j, \pi(i)} x_{j, \pi(i)} \ge \sum_{i \ge k_j} \Delta_j x_{j, \pi(i)} = \Delta_j \sum_{i \ge k_j} x_{j, \pi(i)} \ge \Delta_j (1 - \alpha),
        \end{align*}
        where the second inequality holds since $\Delta_j = \max_{f: \overline{x}_{j, f} > 0} \texttt{dist}_{j, f} = \texttt{dist}_{j, \pi(k_j)} \ge \texttt{dist}_{j, \pi(i)}$ for all $i \ge k_j$ by definition of ordering $\pi$. The third inequality holds since $k_j$ is the first index such that $\sum_{i \le k_j} x_{j, \pi(i)} \ge \alpha$, so that $\sum_{i < k_j} x_{j, \pi(i)} < \alpha$. Therefore, since $\alpha = \frac{\delta + 2}{2(\delta + 1)}$, we have
        \[
            \Delta_j \le \frac{1}{1 - \alpha} \dist_j = \frac{2(\delta + 1)}{\delta} \dist_j \le \frac{4}{\min(1, \delta)} \dist_j.
        \]
        \item Intuitively, loss for a facility increases due to an increase in the operating cost $c_f y_f$, and a decrease in client revenue. Denote by $F_1 = \{f \in F: c_f \overline{y}_f > \sum_{j \in C} r_j \overline{x}_{j, f}\}$ the set of unprofitable facilities in fractional solution $(\overline{x}, \overline{y})$. Then,
        \begin{align*}
            \sum_{f \in F} \overline{\loss_f} &= \sum_{f \in F_1} \overline{\loss_f} = \sum_{f \in F_1} \bigg(c_f \overline{y}_f - \sum_{j \in C} r_j \overline{x}_{j, f}\bigg) \\
            &= \frac{1}{\alpha} \sum_{f \in F_1} c_f y_f - \sum_{j \in C} r_j \sum_{f \in F_1} \overline{x}_{j, f}.
        \end{align*}
        Next, from Lemma \ref{lem: alpha-point-rounding-mass-shift}, we have that $\sum_{f \in F_1} \overline{x}_{j, f} \ge - (1 - \alpha) + \sum_{f \in F_1} \overline{x}_{j, f}$ for all clients $j \in C$, so that the above becomes
        \begin{align*}
            \sum_{f \in F} \overline{\loss_f} &\le \frac{1}{\alpha} \sum_{f \in F_1} c_f y_f - \sum_{j \in C} r_j \sum_{f \in F_1} x_{j, f} + (1 - \alpha) \sum_{j \in C} r_j. 
        \end{align*}
        Next, we note that $c_f y_f \le \loss_f + \sum_{j \in C} r_j x_{j, f}$ since $(x, y)$ satisfies constraint (\ref{eqn: loss-is-more-than-cost-gap}) in (\ref{eqn: price-of-fairness-ip}), so that the above becomes
        \begin{align*}
            \sum_{f \in F} \overline{\loss_f} &\le \frac{1}{\alpha} \left(\sum_{f \in F_1} \loss_f + \sum_{j \in C} r_j \sum_{f \in F_1} x_{j, f}\right) - \sum_{j \in C} r_j \sum_{f \in F_1} x_{j, f} + (1 - \alpha) \sum_{j \in C} r_j \\
            &= \frac{1}{\alpha} \sum_{f \in F_1} \loss_f + \left(\frac{1}{\alpha} - 1\right) \sum_{j \in C} r_j \sum_{f \in F_1} x_{j, f} + (1 - \alpha) \sum_{j \in C} r_j \\
            &\le \frac{1}{\alpha} \cdot \delta \sum_{j \in C} r_j + \left(\frac{1}{\alpha} - 1\right) \sum_{j \in C} r_j + (1 - \alpha) \sum_{j \in C} r_j \\
            &\le \left(\frac{\delta}{\alpha} + 2 \left(\frac{1}{\alpha} - 1\right)\right) \sum_{j \in C} r_j.
        \end{align*}
        The second last inequality follows since $(x, y)$ has subsidy $\le \delta$, and the last inequality follows since $(1 - \alpha) \le (1/\alpha - 1)$ for all $\alpha \in (0, 1)$. Setting $\alpha = \frac{\delta + 2}{2(\delta + 1)}$ gives $\frac{\delta}{\alpha} + 2 \left(\frac{1}{\alpha} - 1 \right) = \frac{\delta + 2}{\alpha} - 2 = 2 \delta$. \qed
\end{enumerate}

\subsection{Proof of Lemma \ref{lem: integral-facility-bounded-loss}}\label{app: integral-facilities-bounded-loss-proof}

    For every core client $j^* \in C^*$,
    \begin{align*}
        c_{f(j^*)} &= c_{f(j^*)} \sum_{f \in \texttt{feas}_{j^*}} \overline{x}_{j^*, f} & \left(\text{constraint} \ (\ref{cons: assign-every-client})\right) \\
        &\le c_{f(j^*)} \sum_{f \in \texttt{feas}_{j^*}} \overline{y}_{f} \ & \left(\text{constraint} \ (\ref{cons: assign-clients-only-to-open-facilities})\right) \\
        &\le \sum_{f \in \texttt{feas}_{j^*}} c_{f} \overline{y}_f. & \left(\text{step} \ (\ref{step: open-cheapest-facility})\right)
    \end{align*}
    Further, the revenue of clients assigned to $f(j^*)$ is
    \begin{align*}
        \sum_{j \in C} x'_{j, f(j^*)} &\ge \sum_{j \in C} r_j \sum_{f \in \feas_{j^*}} \overline{x}_{j, f} & \left(\text{step} \ (\ref{step: client-assignments-integral-facilities})\right) \\
        &= \sum_{f \in \feas_{j^*}} \sum_{j \in C} r_j \overline{x}_{j, f} \\
        &\ge \sum_{f \in \feas_{j^*}} \left(c_f \overline{y}_f - \overline{\loss}_f\right). & \left(\text{constraint} \ (\ref{eqn: loss-is-more-than-cost-gap})\right)
    \end{align*}
    Denote unprofitable facilities in $(x', y')$ by $F_1 = \{f \in F: \loss'_f > 0\}$. Since the only open facilities in $(x', y')$ are $f(j^*), j^* \in C^*$, we get from above that the total loss of $(x', y')$ is
    \begin{align*}
        \sum_{f\in F} \loss'_f &= \sum_{j^*: f(j^*) \in F_1} \left[c_{f(j^*)} - \sum_{j \in C} r_j \overline{x}_{j, f(j^*)}\right] \\
        &\le \sum_{j^*: f(j^*) \in F_1} \left[ \sum_{f \in \feas_{j^*}} \left[c_f \overline{y}_f - (c_f \overline{y}_f - \overline{\loss}_f)\right]\right] \\
        &= \sum_{j^*: f(j^*) \in F_1}  \sum_{f \in \feas_{j^*}} \overline{\loss}_f \le \sum_{j^* \in C^*}  \sum_{f \in \feas_{j^*}} \overline{\loss}_f.
    \end{align*}
    Since sets $\feas_{j^*}, j^* \in C^*$ are disjoint for different core clients $j^*$, we get that this is bounded by $\sum_{f \in F} \overline{\loss}_f$. \qed

\subsection{Proof of Lemma \ref{lem: integral-assignment}}\label{app: integral-assignment}

In this section, we prove Lemma \ref{lem: integral-assignment} that given a $\delta'$-subsidized $\Delta'$-close fractional solution $(x', y')$ with integral $y'$, obtains a $(\delta' + \theta)$-subsidized $\Delta'$-close integral solution $(x'', y')$.

\emph{Proof of Lemma \ref{lem: integral-assignment}}. We prove the guarantee of distances first, and then on the subsidy.
    
\textbf{Distances.} By the lemma, $x''$ only assigns $j$ to some facility $f$ such that $x'_{j, f} > 0$ (i.e., $f$ is in the support of $x'_j$). Therefore, $\texttt{dist}_{j, f} \le \Delta'_j$ by definition.
     
\textbf{Subsidy.} For brevity, let us denote the initial load/revenue of facility $f$ by $r'(f) := \sum_{j \in F_C} x'_{j, f} r_j$, the final load/revenue by $r''(f) := \sum_{j \in F_C} x''_{j, f} r_j$, and the change in revenue by $\rho(f) := r''(f) - r'(f)$. The above then says that $\rho(f) \le r_{j(f)}$ for some $j \in C_f$ for all $f \in F$. By the small revenues assumption, $r_{j(f)} \le \theta c_f$, so that 
\[
    r''(f) - r'(f) = \rho(f) \le \theta c_f.
\]
Denote $F_{\uparrow}$ to be the set of all facilities $f$ whose load/revenue increases, i.e., $\rho(f) \ge 0$, and similarly $F_{\downarrow}$ is the set of all facilities $f$ whose load/revenue decreases. Since the total revenue $r(C) := \sum_{j \in C} r_j$ is constant, the total decrease in revenue across facilities $F_\downarrow$ must come from the total increase in revenue across facilities $F_\uparrow$. Therefore, to track the total decrease in revenue in $F_\downarrow$ (and therefore the increase in total loss), we will track the total increase in revenue across facilities in $F_\uparrow$. Consider a facility $f \in F_\uparrow$. There are three cases:
\begin{enumerate}
    \item $f$ was profitable in $(x', y')$, i.e., we had $c_f \le r'(f)$. In this case, the contribution of $f$ to the increase in revenue is
    \[
        \rho(f) \le \theta c_f \le \theta r'(f) = \theta (r''(f) - \rho(f)) \le \theta r''(f).
    \]
    \item $f$ was unprofitable in $(x', y')$ and loss $\loss'_f$ of $f$ in $(x', y')$ was at least $\rho(f)$, i.e., $\loss'_f \ge \rho(f)$. In this case, the new loss of $f$ in $(x'', f')$ is $\loss''_f = \loss'_f - \rho(f)$. That is, loss decreases by $\rho(f)$. Then, any decrease in revenue in $F_\downarrow$ due to $\rho(f)$ is compensated by this decrease in loss.
    \item $f$ was unprofitable in $(x', y')$ and loss $\loss'_f$ of $f$ in $(x', y')$ was at most $\rho(f)$, i.e., $\loss'_f \le \rho(f)$. In this case,
    \begin{align*}
        \rho(f) &\le \theta c_f = \theta\left(\loss'_f + r'(f)\right) \le \theta\left(\rho(f) + r'(f)\right) = \theta r''(f).
    \end{align*}
\end{enumerate}
The total increase in revenue in cases 1 and 3 combined is therefore at most
\[
    \theta \sum_{f \in F_\uparrow} r''(f) = \theta  \sum_{f \in F_\uparrow} \sum_{j \in C} x''_{j, f} r_j = \theta \sum_{j \in C} r_j \sum_{f \in F_\uparrow} x''_{j, f} \le \theta \sum_{j \in C} r_j =  \theta r(C).
\]
That is, the total decrease in loss overall when rounding from $(x', y')$ to $(x'', y')$ is upper bounded by $\theta r(C)$, thus increasing the subsidy by at most $\theta$. \qed

\section{$k$-Clustering and Uncapacitated Facility Location}\label{app: clustering-and-ufl}

In this section, we show that the $k$-Clustering and Uncapacitated Facility Location (UFL) problems are special cases of the Fair Subsidized Facility Location (FSFL) problem.

The $k$-Clustering problem is similar to FSFL except we are only given the metric space $(C \cup F, \texttt{dist})$, the objective $g$, and a bound $k$ on the number of facilities as input and must return a solution $(F', \Pi)$ with $|F'| \le k$ that minimizes $g$. Examples of $k$-Clustering include the $k$-median problem, where $g$ is the sum of client distances to corresponding assigned facilities, the $k$-mean problem, where $g$ is the $L_2$ norm of the client distances, and the $k$-center problem, where $g$ is the maximum client distance.

In UFL, we are only given the metric space $(C \cup F, \texttt{dist})$, operating costs $c_f > 0$ for $f \in F$, and the distance objective function $g$. Given a solution $(F', \Pi)$ with client distances $\dist_j, j \in C$, its objective value is the sum $g(\dist) + \sum_{f \in F'} c_f$ of operating costs of open facilities and the distance objective function value.

We reduce both these problems to FSFL:
\begin{theorem}\label{thm: clustering-and-ufl}
    $k$-Clustering and UFL are special cases of FSFL.
\end{theorem}

    \emph{Proof.} We reduce the following \emph{budgeted facility location problem} to FSFL; it is standard to reduce both $k$-Clustering and UFL to this budgeted problem. In this budgeted problem, we are given (i) the metric space $(C \cup F, \texttt{dist})$, (ii) operating costs $c_f > 0$ for $f \in F$, (iii) a budget $B$ on operating cost of open facilities, and (iv) the distance objective function $g$. A solution $(F', \Pi)$ is feasible if $\sum_{f \in F'} c_f \le B$ and its objective value is $g(\dist)$ where $\dist$ is the vector of client distances to assigned facilities. Clearly, $k$-Clustering is a special case of this problem with $c_f = 1$ for all $f$ and $B = k$. UFL can be reduced to this problem by `guessing' the operating cost $B$ of open facilities in the optimal solution and minimizing $g$ while fixing this operating cost.

    Given an instance of the budgeted problem, assume (after possibly scaling operating costs, budget $B$, and distances $\texttt{dist}$) without loss of generality that $\min_{f \in F} c_f = 1$. We construct an instance of FSFL as follows: metric space $(C \cup F, \texttt{dist})$, operating costs $c$, and distance objective function $g$ stay the same. We define revenue $r_j := \frac{1}{|C|}$ for each client $j \in C$, and set $\delta = B - 1$.

    First, we show that a feasible solution $(F', \Pi)$ to the budgeted problem is also feasible for the FSFL instance. It is sufficient to show that it is $\delta$-subsidized, that is, the total loss of unprofitable facilities is at most $\delta$ times the total revenue $\sum_{j \in C} r_j$ of all clients. Note that since $\min_f c_f = 1$ and $\sum_{j \in C} r_j = |C| \times \frac{1}{|C|} = 1$, we get that all open facilities are unprofitable, so that $\loss_f = c_f - \sum_{j: \Pi(j) = f} r_j$ for all $f \in F'$. Therefore, the total loss of facilities is
    \[
        \sum_{f \in F'} \loss_f = \sum_{f \in F} \left(c_f - \sum_{j: \Pi(j) = f} r_j\right) = \left(\sum_{f \in F} c_f\right) - \sum_{j \in C} r_j \le B - 1 = \delta = \delta \sum_{j \in C} r_j,
    \]
    where the inequality follows since $(F', \Pi)$ is feasible for the budgeted problem.

    Conversely, we show that a feasible solution $(F', \Pi)$ to FSFL is feasible for the budgeted problem: as before, all open facilities are unprofitable, so that
    \[
        B - 1 = \delta = \delta \sum_{j \in C} r_j \ge \sum_{f \in F'} \loss_f = \sum_{f \in F} \left(c_f - \sum_{j: \Pi(j) = f} r_j\right) = \left(\sum_{f \in F} c_f\right) - \sum_{j \in C} r_j = \left(\sum_{f \in F} c_f\right) - 1.
    \]
    Therefore, $\sum_{f \in F'} c_f \le B$. Observing that the objective values stay exactly the same finishes the proof. \qed
\section{Hardness Results for FSFL}\label{app: hardness}

In this section, we prove Theorem \ref{thm: hardness}, which states that FSFL is hard to approximate within constant factors (assuming P $\neq $ NP), and that even checking whether a solution is feasible for the problem may be NP-hard. This implies that bicriteria approximation oracles are unavoidable for the problem.

\hardness*

\begin{figure}
    \centering
    \includegraphics[width=0.7\columnwidth]{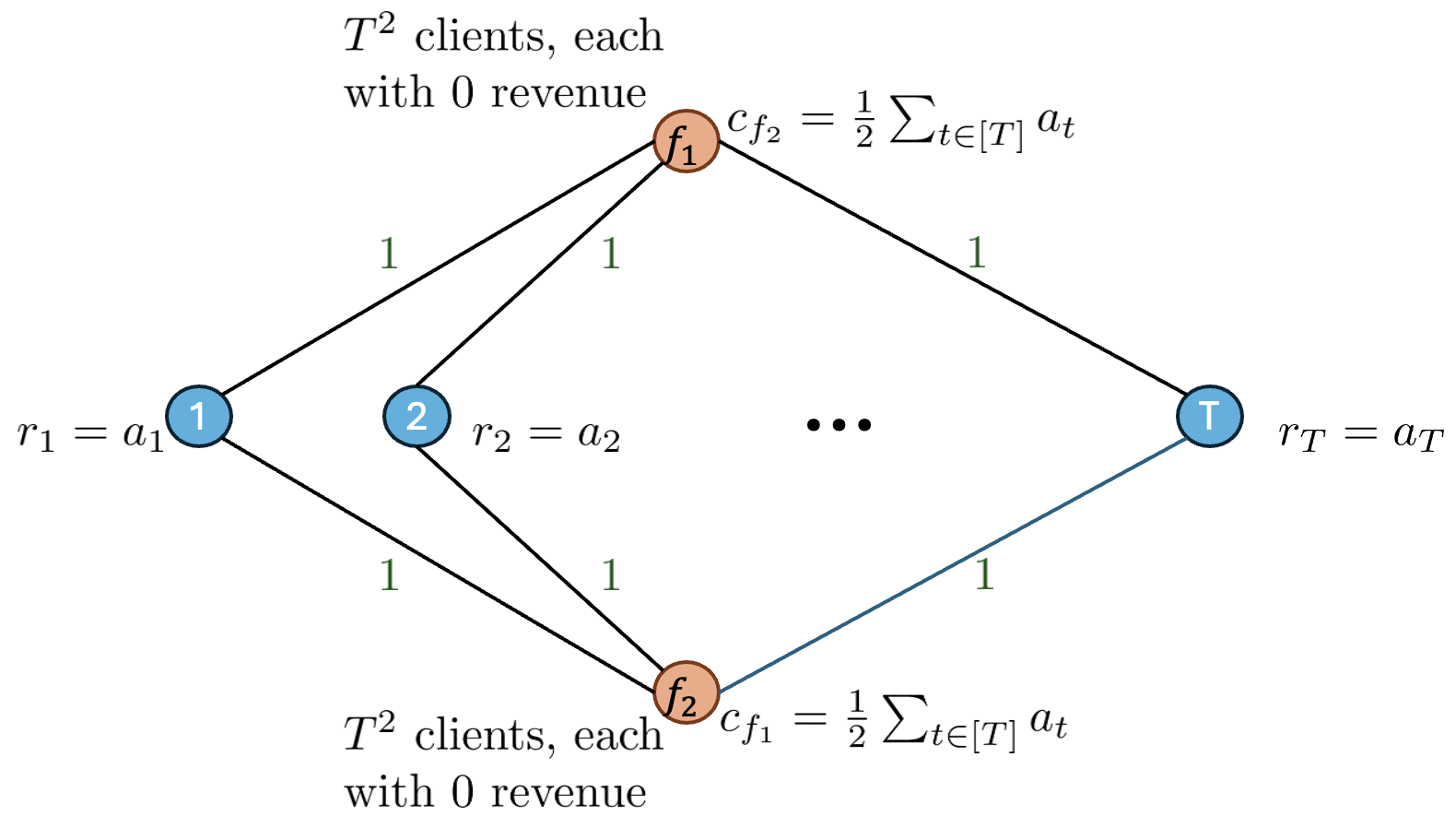}
    \caption{The FSFL instance used in proof of Theorem \ref{thm: hardness}}
    \label{fig: hardness-subset-sum-reduction}
\end{figure}

\emph{Proof.}
    We show that solving FSFL to within any constant factor implies a polynomial-time algorithm for the Subset Sum problem, which in turn implies P = NP. Recall that in the Subset Sum problem, we are given a set $A = \{a_1, \ldots, a_T\}$ of positive integers and need to determine whether we can partition $A$ into two subsets with equal sum.

    Consider a metric space with $|C| = T + 2T^2$ clients and $|F| = 2$ facilities. Each of the two facilities $f_1$ and $f_2$ has opening cost $c = c_{f_1} = c_{f_2} = \frac{1}{2} \sum_{t \in [T]} a_t$. There are also $T^2$ clients, each with revenue $0$ at the location of each of the two facilities. The remaining $T$ clients have revenues $a_1, \ldots, a_T$ respectively, and each of them is at a distance of $1$ from each facility; see Figure \ref{fig: hardness-subset-sum-reduction}. We set the subsidy parameter $\delta = 0$, and the objective function $g: \R^{C} \to \R$ is the $L_1$ norm, i.e., we seek to minimize the sum of distances of clients to their assigned facilities.

    Consider an algorithm $\mathcal{A}$ that is a constant-factor approximation for FSFL. We construct an algorithm for subset-set as follows: (i) Run $\mathcal{A}$ on the above instance. (ii) If the solution $(F', \Pi)$ returned by $\mathcal{A}$ opens both facilities, declare ``Yes'' (i.e., there is a subset $S \subseteq A$ whose elements sum exactly $\frac{1}{2}\sum_{t \in T} a_t$). (iii) If $(F', \Pi)$ opens only one $|F'| = 1$ facility, declare ``No''.

    The key observation is that the following statements are equivalent:
    \begin{enumerate}
        \item $A$ can be partitioned into two subsets with equal sums.
        \item Solution returned by $\mathcal{A}$ has objective value $O(T)$
        \item Facility set $F'$ output by algorithm $\mathcal{A}$ opens both facilities, i.e., $|F'| = 2$. 
    \end{enumerate}

    It is easy to see that (2) implies (3): if one of the facilities is not open, all $T^2$ clients at that location travel distance $2$ to the other facility so that the objective value is $\ge 2T^2$. Next, if (3) holds, then since this solution is feasible for FSFL, the total loss must be $\delta \times R = 0 \times \sum_{j \in C} r_j = 0$. That is, we must have $\sum_{j: \Pi(j) = f_1} r_j \ge c_{f_1} = \frac{1}{2} \sum_{t \in [T]} a_t$ and similarly $\sum_{j: \Pi(j) = f_2} r_j \ge c_{f_2} = \frac{1}{2} \sum_{t \in [T]} a_t$. Since $\sum_t a_t = \sum_j r_j = \sum_{j: \Pi(j) = f_1} r_j + \sum_{j: \Pi(j) = f_2} r_j$, we get a partition of $A$ into two sets of equal sum.

    Finally, we show that (1) implies (2): let $(S, A \setminus S)$ be the partition of $A$ with equals sums. Consider the following solution to FSFL: open both facilities, and assign each $a_t \in S$ to $f_1$ and each $a_t \in A \setminus S$ to $f_2$. Similar to the argument above, this solution is feasible for the problem. Further, it has objective value $T$, which is the best possible. Since $\mathcal{A}$ is an $O(1)$-approximation for FSFL, its objective value must be $O(T)$. This concludes the proof of Part 1.

    This equivalence also implies that checking feasibility is NP-hard: indeed, the solution with both facilities open is feasible if and only if $A$ can be partitioned into two subsets with equal sums. \qed

\section{Lower Bound on Portfolio Size for FSFL}\label{app: fsfl-portfolio-lower-bound}

In this Section, we prove Theorem \ref{thm: portfolios-ffl-lower-bound}, which gives a lower bound of $\Omega(\log t)$ on portfolios of $L_p$ norms for FSFL with $t$ client groups.

\emph{Proof.} For all large enough $t$ and for all constant $\alpha > 1$, we give an instance of FSFL where any $\alpha$-approximate portfolio size must be $\frac{1}{4} \log_{2\alpha} t = \Omega(\log_{2\alpha} t)$. Fix $t$. Each client will be in their own unique group, so the number of clients $|C| = t$. Denote $\gamma = 2\alpha$, and denote $L$ to be the highest integer that satisfies $1 + L(\gamma^{2L} + 1) = t$; then $t \ge \frac{1}{4}\log_{\gamma} L$.

Our metric space is a star graph with central vertex $a_0$ and leaf vertices $a_1, \ldots, a_L$ with unit distances between $a_0$ and $a_i$ for all $i \in [L]$. A facility can be opened at any vertex (including the central vertex $a_0$) with operating cost $c = 1$.

The clients are specified as follows. While each client is in their own unique group, the group distances will be weighted, i.e., client in group $s \in [t]$ will have a weight $\mu_s$ so that the $L_p$ norm objective for traveled distances $d_s, s \in [t]$ is $\left(\sum_{s \in [t]} (\mu_s d_s)^p\right)^{1/p}$.
\begin{itemize}
    \item There is a single client at $a_0$ with weight $\gamma^{L^2}$ and revenue $0$.
    \item At $a_i$, there are $\gamma^{2L} + 1$ clients, each with revenue $r = \frac{1}{\gamma^{2L} + 1}$. The first client has weight $\gamma^i$ while the other $\gamma^{2L}$ clients have weight $\gamma^{-i}$ each.
\end{itemize}

Set subsidy $\delta = \frac{1}{2L}$. Note that since the sum of revenues is $r \times (\gamma^{2L} + 1) \times L = L$ and since each of the $L + 1$ facilities has an operating cost $c = 1$, we can open at most $L$ facilities, i.e., a facility must not be open at some $a_i, i \in [0, L]$, and all clients at $a_i$ must travel a unit distance to some other open facility.

If a facility is not open at $a_0$, the weighted group distance vector is $d^{(0)} = (\gamma^{L^2}, 0, \ldots, 0)$. If a facility is not open at $a_i$ for $i \in [L]$, the weighted group distance vector is $d^{(i)} = (\gamma^i, \underbrace{\gamma^{-i}, \ldots, \gamma^{-i}}_{\gamma^{2L}}, 0, \ldots, 0)$. For any $p \ge 1$, the $L_p$ norm of these vectors is
\[
    \|d^{(i)}\|_p = \begin{cases}
        \gamma^{L^2} & \text{if} \ i = 0, \\
        (\gamma^{ip} + \gamma^{2L - ip})^{1/p} & \text{if} \ i \in [L].
    \end{cases}
\]
Consider $p = L/j$ for $j \in \{1, \ldots, L\}$. Then $\|d^{(i)}\|_p$ is minimized at $i = j$ with value $2^{j/L} \cdot \gamma^j \le 2 \cdot \gamma^j$. For all $i > j$, the first term dominates, and $\|d^{(i)}\|_p > \gamma^i \ge (2\alpha) \cdot \gamma^j = \alpha \cdot (2\gamma^j)$. For all $i < j$, the second term dominates, and $\|d^{(i)}\|_p > \gamma^{2L - i} \ge 2\alpha \cdot \gamma^j$. That is, the only $\alpha$-approximate solution for the $L_p$ norm objective is to not open the facility at $a_j$.

Thus, any $\alpha$-approximate portfolio for $L_p$ norms $\left\{\frac{L}{1}, \frac{L}{2}, \ldots, \frac{L}{L}\right\}$ must contain $L$ distinct solutions. \qed

\section{Gap Between Portfolio Sizes for Top-$\ell$ Norms and $L_p$ Norms}\label{app: portfolios-are-not-transferable}

In this section, we show that portfolios for one class of interpolating functions may not be portfolios for another class of interpolating functions. Specifically, we give a feasible set $\mathcal{D} \subseteq$ and base functions $h_1, \ldots, h_N: \mathcal{D} \to \R_{> 0}$ where the class of top-$\ell$ norms admits an \emph{optimal} portfolio of size $2$ but any $O(1)$-approximate portfolio for $L_p$ norms must have size $\simeq (\log N)^{1/3}$. This shows that large gaps between portfolio sizes for different norms are possible. Compare this with what happens for portfolios of size $1$ or simultaneous approximations, where \cite{goel_simultaneous_2006} show that $\alpha$-approximate simultaneous approximations for top-$\ell$ norms are also $\alpha$-approximate simultaneous approximations for $L_p$ norms.

\begin{lemma}
    For all large enough $N \in \Z_{> 0}$, there exists a set $\mathcal{D}$ and base functions $h_1, \ldots, h_N: \mathcal{D} \to \R_{> 0}$ such that
    \begin{enumerate}
        \item There is an optimal portfolio $P$ of size $2$ for all top-$\ell$ norms, and
        \item Any $O(1)$-approximate portfolio $P'$ for all $L_p$ norms must have size $\Omega\left(\left(\frac{\log 
        N}{\log\log N}\right)^{1/3}\right)$.
    \end{enumerate}
\end{lemma}

\emph{Proof.}
    We will construct a set $\mathcal{D} \subseteq \R^{N}$, and our $N$ base functions will be $h_i(x) = x_i$ for all $x \in \mathcal{D}$.

    Let $S = S(N)$ be a super-constant that we will fix later, and $L$ be the largest integer such that $S^{L^2} \le n$. Then $L = \Omega(\sqrt{\log_S N})$. For $s \in [1, L]$, define the vectors $v(s) \in \R_{\ge 0}^N$ as:
    \[
        v(s) = (\underbrace{S^{-2s}, \ldots, S^{-2s}}_{S^{s^2}}, 0, \ldots, 0).
    \]

    Let $\mathcal{D} = \{v(1), \ldots, v(L)\}$. We will prove Part 1 of the lemma statement first. Specifically, we will show that either $v(1)$ or $v(L)$ is optimum for all top-$\ell$ norms, so that $P = \{v(1), v(L)\}$ is an optimal portfolio for all top-$\ell$ norms. We have
    \[
        \left( \text{top-}\ell \: \text{norm of}\: v(s)\right) = \begin{cases}
            \ell S^{-2s} & \text{if}\: \ell \le S^{s^2},  \\
            S^{s^2 - 2s} & \text{if}\: \ell >  S^{s^2}.
        \end{cases}
    \]
    Fix $\ell$. For $s$ such that $S^{s^2} < \ell$, $\left( \text{top-}\ell \: \text{norm of}\: v(s)\right) = S^{s^2 - 2s}$ increases as $s$ increases since $s \ge 1$. For $s$ such that $S^{s^2} > \ell$, $\left( \text{top-}\ell \: \text{norm of}\: v(s)\right) = \ell S^{-2s}$ decreases as $s$ increases. Therefore, for each top-$\ell$ norm, either $v(1)$ or $v(L)$ is optimum. This proves Part 1 of the theorem statement.

    We move to Part 2 of the theorem statement. Consider $L_p$ norms for $p \in \{1, \ldots, L\}$. Then we claim that for appropriate choice of $S = S(N)$, (1) for all $p \in [1, L]$, ${\arg\min}_{v(s)} \|v(s)\|_{p} = v(p)$. That is, vector $v(p)$ has the minimum norm $\|\cdot\|_{p}$ among all vectors in $\mathcal{D}$, and (2) for all $p \in [1, L]$ and $s \neq p$, $v(s)$ is not an $O(1)$-approximation for minimizing $\|\cdot\|_{p}$. Together, the two claims imply that any $O(1)$-approximate portfolio for $L_p$ norms, $p \in \{1, \ldots, L\}$ must contain each of $v(1), \ldots, v(L)$. Note first
    \begin{equation}
        \|v(s)\|_{p} = \left(S^{s^2} \cdot S^{2ps}\right)^{1/p} = S^{\frac{s^2}{p} - 2s}.
    \end{equation}
    To show that this is minimum at $s = p$, consider $f(x) = \frac{x^2}{p} - 2x$. It attains its minimum at $x = p$. Since $S > 1$, this implies that ${\arg\min}_{v(s)} \|v(s)\|_p = v(p)$, and the minimum is $S^{-p}$.

    Further, for any $s \neq p$, say $s = p + \theta$ for non-zero integer $\theta$, we have $\log_S \|v(s)\|_{p} = \frac{(p + \theta)^2}{p} - 2 (p + \theta) = (p + \theta)\left(1 + \frac{\theta}{p} - 2\right) = \frac{\theta^2 - p^2}{p} = \frac{\theta^2}{p} - p = \frac{\theta^2}{p} + \log_S \|v(p)\|_p$.
    Therefore, $\frac{\|v(s)\|_p}{\|v(p)\|_p} \ge S^{\frac{\theta^2}{p}} \ge S^{\frac{1}{p}} = \exp((\log S)/p)$ Since $p \le L$, this is at least $\exp(\frac{\log S}{L})$.
    We choose $S$ such that $\log S = (\log N)^{1/3} (\log\log N)^{2/3}$. Then $L = \Theta(\sqrt{\log_S N}) = \Theta\left(\sqrt{\frac{\log N}{\log S}}\right) = \Theta\left(\left(\frac{\log N}{\log\log N}\right)^{\frac{1}{3}}\right)$. Therefore, $\log \left(\frac{\|v(s)\|_p}{\|v(p)\|_p}\right) \ge \frac{\log S}{L} = \Theta\left(\log\log N\right)$. That is, $\|v(s)\|_p = \omega(\|v(p)\|_p)$. This proves claim 2. Lastly, note that the size of portfolio $\{v(1), \ldots, v(L)\}$ for $L_p$ norms is $L = \Theta\left(\left(\frac{\log N}{\log\log N}\right)^{1/3}\right)$.

\section{Further Details on Experiments}\label{app: further-details-on-experiments}

We make the following assumptions and modeling choices for our experiments. Since Census blockgroup is the most granular level at which Census data is provided, we assume that all people in each Census blockgroup are located at the geometric center of the blockgroup. All our computations assume straight-line distances. We assume that a new facility can be located at the center of any blockgroup. A facility cannot be opened outside Mississippi, although we allow people to travel to existing facilities outside Mississippi if those are closer. For simplicity, all operating costs are uniformly set to $\$2500$ while client revenues are set to $\$0.10$ for each person above the poverty line and $\$0.05$ for each person below the poverty line. These parameters can be varied in our model.

\end{document}